\documentclass[aps,prd,twocolumn,superscriptaddress,groupedaddress,nofootinbib,floatfix]{revtex4}

\usepackage{graphicx}
\usepackage{dcolumn}
\usepackage{bm}

\usepackage{amsmath}	
\usepackage{amssymb}	
\usepackage{epsfig,amsmath,natbib}
\usepackage{color,subfigure}
\usepackage{xcolor}
\usepackage{mathrsfs,amssymb,amstext}
\usepackage{ulem}
\usepackage{url}
\usepackage{adjustbox}
\usepackage{threeparttable}
\usepackage{longtable}
\usepackage{aas_macros}

\newcommand{\Mpc}{\mathrm{~km~s^{-1}~Mpc^{-1}}}
\newcommand{\kms}{\mathrm{~km~s^{-1}}}

\begin{document}


\title{Effects of a central dark matter core on time-delay cosmography with galaxy clusters} %

\author{Yuting Liu\footnote{{yutingl@imu.edu.cn}}}
\affiliation{Institute of Astronomy and Physics,  Inner Mongolia University, Hohhot 010021, China }
\affiliation{School of Physical Science and Technology, Inner Mongolia University, Hohhot 010021, China}

\author{Masamune Oguri\footnote{masamune.oguri@chiba-u.jp}} 
\affiliation{Center for Frontier Science, Chiba University,  Chiba 263-8522, Japan}
\affiliation{Department of Physics, Graduate School of Science, Chiba University, Chiba 263-8522, Japan}

\date{\today}

\begin{abstract}
The central dark matter distribution probes the nature of dark matter and baryon physics, yet its impact on time-delay cosmography is not fully explored. We update previous analyses of the cluster-lensed quasar SDSS J1004+4112 and the supernova Refsdal in MACS J1149.5+2223 by replacing the main halo density profile from an elliptical Navarro-Frenk-White (NFW) profile to a cored elliptical NFW profile and investigating the impact of the central dark matter core on the Hubble constant $H_0$. We jointly infer $H_0$ and the dimensionless core size parameter $\beta=r_{\rm c}/r_{\rm s}$, while the Brightest Cluster Galaxy (BCG) is either tied to the scaling relation of cluster member galaxies or modeled separately, with its mass left free or assigned a Gaussian prior from the stellar mass estimate for a Chabrier or Salpeter initial mass function (IMF). For SDSS J1004+4112, constraints on both $\beta$ and $H_0$ strongly depend on the treatment of the BCG. A large core is implied when no prior is added to the BCG stellar mass, while a reasonable assumption on the BCG mass prefers a nearly cuspy halo with $\beta\sim 10^{-3}$. For MACS J1149.5+2223, a large core of $\beta\sim 0.5$ is obtained with no prior, while a modest core of $\beta=0.0410^{+0.0865}_{-0.0276}$ is preferred when the Salpeter IMF is assumed. The constraint on $H_0$ from MACS J1149.5+2223 is stable against the BCG treatment, supporting its robustness. For the Salpeter-prior case, we obtain $H_0=65.9^{+3.5}_{-3.4}\mathrm{~km~s^{-1}~Mpc^{-1}}$, and different BCG assumptions change the median $H_0$ by less than $1\sigma$. The core size constraints are translated into the self-interacting dark matter cross section, giving a 95\% upper limit of $\sigma/m<0.035\,\mathrm{cm}^{2}\,\mathrm{g}^{-1}$ from SDSS J1004+4112 and the 68\% range of $\sigma/m=0.071^{+0.15}_{-0.045}\,\mathrm{cm}^{2}\,\mathrm{g}^{-1}$ from MACS J1149.5+2223.

\end{abstract}


\maketitle


\section{Introduction}\label{sec:intro}

Measurements of the present expansion rate of the Universe remain 
discrepant.  Within a flat $\Lambda$-dominated Cold Dark Matter 
($\Lambda$CDM) model, the cosmic microwave background anisotropies
measured by the Planck satellite imply
$H_0=67.4\pm0.5\Mpc$ \cite{2020A&A...641A...6P}, whereas the local
distance ladder calibrated with Cepheids gives
$H_0=73.04\pm1.04\Mpc$ \cite{2022ApJ...934L...7R}, representing 
a difference of nearly $5\sigma$.  Establishing
whether this difference reflects residual systematics or physics
beyond the standard cosmological model requires measurements with
independent distance calibrations and different astrophysical
uncertainties.

Time-delay cosmography provides such an independent measurement
\cite{1964MNRAS.128..307R,2024SSRv..220...48B}.  
For multiple images of a time-variable source, the observed time 
delay is the product of the time-delay 
distance and the Fermat-potential difference predicted by 
the lens model. Since the time-delay distance scales as 
$H_0^{-1}$, an accurate inference of $H_0$ requires both a precise 
measurement of the time delay and an accurate reconstruction of 
the lens potential. The main limitation is no longer 
the measurement of time delays, but the freedom in the radial 
mass profile and the related mass-sheet degeneracy 
\cite{1985ApJ...289L...1F}.  Recent galaxy-scale analyses address 
this problem by combining lensing with stellar kinematics and 
line-of-sight information.  The TDCOSMO-2025 analysis of eight lensed 
quasars, using new JWST, Keck, and VLT spectroscopy and allowing 
conservative freedom in the mass profile, indicates 
$H_0=71.6^{+3.9}_{-3.3}\Mpc$ in a flat $\Lambda$CDM model
\cite{2025A&A...704A..63T}.  This result illustrates both the maturity
of the method and the sensitivity of cosmographic inferences to the
assumed mass distribution.

Galaxy clusters provide a complementary method for time-delay
cosmography.  Their lens potentials are constrained not only by the
time-variable source but also by multiple images of many other background
galaxies, often spanning a wide range of source redshifts and
projected radii.  At the same time, cluster lenses contain several
cluster-scale halos, a population of galaxy-scale perturbers, the
brightest cluster galaxy (BCG), and structure along the line-of-sight.
Their additional constraining power is therefore accompanied by a
more complex mass-modeling problem.  For the cluster-lensed quasar
SDSS J1004+4112, an analysis of 16 parametric mass models yields
$H_0=67.5^{+14.5}_{-8.9}\Mpc$, with the uncertainty dominated by the
variation among lens models \cite{2023PhRvD.108h3532L} (see also \cite{2023ApJ...959..134N}). A comparable
spread among independent reconstructions of the same cluster is
noted by the Frontier Fields lens-modeling comparison project
\cite{2017MNRAS.472.3177M}.  For SN Refsdal in MACS J1149.5+2223
\cite{2015Sci...347.1123K,2016ApJ...819L...8K}, the measured
time delay of a reappeared image yields
$H_0=66.6^{+4.1}_{-3.3}\Mpc$ from two models that are most consistent 
with the observations \cite{2023Sci...380.1322K} 
(see also \cite{2024A&A...684L..23G}).  Subsequent work
using a broader set of mass models finds
$H_0=70.0^{+4.7}_{-4.9}\Mpc$ \cite{2025PhRvD.111l3506L}.  With new
VLT/MUSE and JWST spectroscopic redshifts, the latest analysis finds
$H_0=66.0\pm4.3\Mpc$, showing only a modest variation among four
representative mass models \cite{2025PhRvD.112l3526L}.  These studies
illustrate that cluster lensing can deliver a competitive time-delay 
distance when many spectroscopically confirmed multiple images are 
available, but they also leave open which aspects of the central 
mass distribution drive the remaining model dependence.

One of the major uncertainties of the dark matter density profile 
comes from the inner density profile, which is sensitive to the
nature of dark matter as well as baryon physics. Collisionless 
dark-matter-only simulations predict a cuspy Navarro-Frenk-White 
(NFW) density profile \cite{1997ApJ...490..493N}, while baryonic 
processes and dark matter self-interactions can modify the central 
density.  Recent analyses of weak lensing \cite{2026OJAp....961580F} 
and kinematics \cite{2026arXiv260916740L,2026arXiv260919132W} data 
suggest that large dark matter cores are ubiquitous among massive galaxies.
Observationally, however, the existence and size of 
a dark matter core for a cluster-scale dark matter remains
model- and system-dependent, because strong-lensing constraints on a
core are sensitive to the adopted radial profile, halo ellipticity,
central images, and the stellar mass assigned to the Brightest 
Cluster Galaxy (BCG).  Joint analyses of strong and weak lensing 
and the stellar kinematics of BCGs suggest dark matter profiles 
shallower than NFW in some relaxed
clusters \cite{2013ApJ...765...24N,2013ApJ...765...25N}, although the
result is not universal.  A dedicated analysis of Abell~611
constrains the core radius to less than approximately $4$~kpc after
imposing a stellar mass-to-light prior, while also demonstrating that
profile mismatch can generate spurious core constraints
\cite{2019MNRAS.487.1905A}.  A study of eight relaxed clusters places
stringent limits on the dark matter self-interaction cross section from
their inferred core sizes \cite{2022MNRAS.510...54A} (see also \cite{2026PhRvD.113f3531O}).  Conversely,
recent JWST-based lens modeling of AS1063 has reported a much larger
central dark matter core, while emphasizing that the interpretation in
terms of self-interacting dark matter is not unique
\cite{2026A&A...711A.275D}.  Among these modeling choices, the stellar
mass of the BCG is the one that acts on the same central region as the
core itself.

A central core is also directly relevant to time-delay cosmography.  Image
positions primarily constrain derivatives of the lens potential near
the images, whereas time delays depend on differences in the Fermat
potential itself.  Models with different inner density profiles can
therefore reproduce similar image configurations but predict different
time delays, or equivalently different values of $H_0$ for a fixed
observed delay
\cite{2002ApJ...578...25K,2003MNRAS.338L..25O}.  Allowing a finite core
is thus a physically motivated way of opening up this radial freedom.
It is not simply a mass-sheet transformation: a mass sheet rescales the
convergence globally and leaves the image configuration invariant,
whereas a core alters the profile locally, in the region where the
stellar mass of the BCG also contributes.  A core is therefore, in
principle, accessible to independent constraints on the stellar mass.
In a cluster core, this
freedom is coupled to the decomposition between the dark matter halo
and the BCG.  Increasing the stellar mass of the BCG can compensate,
at least partly, for a lower central dark matter density.  It is
therefore necessary to vary the dark matter core and the treatment of
the BCG together rather than interpreting a cored fit in isolation.

In this paper, we examine how a finite central dark matter core affects
time-delay cosmography with the cluster-lensed quasar SDSS J1004+4112
and SN Refsdal in MACS J1149.5+2223, the two cluster-scale time-delay
systems that combine measured delays with the largest sets of
spectroscopically confirmed multiple images and for which detailed
reference mass models are available. Another advantage of these two
clusters is the presence of multiple images near centers of the 
clusters, which is expected to help improve constraints on the inner
dark matter profile (see e.g., \cite{2025MNRAS.541.2341C}). 
Previous analyses of both systems fix the main halo to an NFW 
profile and tied the BCG to the scaling relations of the other 
member galaxies, and to our knowledge, the central core size 
and $H_0$ have not been inferred jointly for either cluster.  
We replace the main NFW halo in each reference model with a
cored NFW profile and infer its core size together with $H_0$.
For each cluster, we compare models in which the
BCG follows the scaling relation of the other member galaxies with
models in which it is represented by a separate galaxy component.
For the latter, we further examine the effect of BCG stellar-mass
priors.  This setup quantifies the relation between the inferred dark
matter core, the stellar mass of the BCG, and the cosmographic 
result. 

The paper is organized as follows.  Section~\ref{sec:data} describes the lensing
data and mass models.  Section~\ref{sec:results} presents 
constraints on the core
size and $H_0$ and examines their dependence on the BCG treatment.
Section~\ref{sec:conclusions} summarizes our conclusion.
Appendix~\ref{app:cnfw} describes the cored elliptical NFW profile and
its implementation, and Appendix~\ref{app:bcg_mass} summarizes our stellar-mass estimates of the BCGs from photometry. Throughout the paper the dimensionless Hubble constant 
$h$ defined by $H_0=100\,h\,\Mpc$ is used.

\section{Data and Mass Modeling}\label{sec:data}

\subsection{Data}

We analyze two cluster-scale time-delay lens systems, the
cluster-lensed quasar SDSS J1004+4112 and the cluster-lensed supernova
Refsdal in MACS J1149.5+2223. Both these systems provide time-delay
measurements together with multiple images of background galaxies at
several different source redshifts, allowing the cluster mass 
distributions to be constrained over a broad range of projected radii.

\subsubsection{SDSS J1004+4112}

SDSS J1004+4112 consists of a galaxy cluster at $z_{\rm l}=0.68$ and a
five-image quasar at $z_{\rm s}=1.734$ \cite{2003Natur.426..810I}.  
The maximum separation between the quasar images is 
$14.62^{\prime\prime}$.  We adopt the observational
constraints compiled in Ref.~\cite{2023PhRvD.108h3532L}, which are based
primarily on the HST Advanced Camera for
Surveys imaging and are summarized in the lens-modeling analysis of
Ref.~\cite{2010PASJ...62.1017O}.  In addition to the positions of the
five quasar images, the data set contains seven multiply imaged knots
in three background galaxies.  The three galaxies lie at
$z_{\rm s}=3.33$, $2.74$, and $3.28$. The first is represented by three
knots with five images each, whereas each of the other two is
represented by two knots with three images. The central fifth image of 
the lensed quasar, which is expected to play an important role in 
constraining the central mass distribution, is spectroscopically 
confirmed \cite{2005PASJ...57L...7I}.

The time-delay constraints, measured relative to quasar image C, are
$\Delta t_{\rm AC}=825.99\pm2.10$ days,
$\Delta t_{\rm BC}=781.92\pm2.20$ days, and
$\Delta t_{\rm DC}=2456.99\pm5.55$ days \cite{2022ApJ...937...34M}.  
We also use the magnitude differences of images B--E relative to image A,
$\Delta m=(0.35\pm0.30,\,0.87\pm0.30,\,1.50\pm0.30,\,
6.30\pm0.80)$, following Ref.~\cite{2023PhRvD.108h3532L}.  The cluster
galaxy catalog contains 90 members selected from the red sequence in
the HST/ACS F435W and F814W images.  Their positions, ellipticities,
position angles, and F814W luminosities are used to specify the galaxy
component of the lens model.

\subsubsection{MACS J1149.5+2223}

MACS J1149.5+2223 \cite{2001ApJ...553..668E} is a galaxy cluster 
at $z_{\rm l}=0.541$ that lenses the host galaxy of SN Refsdal at 
$z_{\rm s}=1.488$.  Four images of the supernova were discovered 
in an Einstein-cross configuration around a cluster member galaxy, 
and a fifth image subsequently appeared at the
position predicted by cluster lens models
\cite{2015Sci...347.1123K,2016ApJ...819L...8K}.  We use the updated
strong-lensing data set assembled in
Ref.~\cite{2025PhRvD.112l3526L}.  It comprises 114 multiple images from
37 background systems, of which 28 have spectroscopic redshifts.  The
constraints include the five images of SN Refsdal, 30 images of
resolved knots in its host galaxy, and 79 images of other background
sources. Some of multiply imaged knots of the host galaxy of the 
supernova Refsdal are located very near the center of the cluster 
(see e.g. \cite{2025PhRvD.111l3506L}), which help constrain the 
central mass distribution of this cluster tightly.

The updated catalog incorporates spectroscopic data from a 5.5-hour
VLT/MUSE observation of the northern part of the cluster
\cite{2024A&A...689A..42S}, together with JWST NIRCam imaging, NIRISS
slitless spectroscopy, and NIRSpec prism spectroscopy from the
CANUCS and JWST in Technicolor programs
\cite{2026ApJS..282....3S}.  These observations add spectroscopic
redshifts, particularly for high-redshift sources, and improve the
strong-lensing constraints in the northern region.  For SN Refsdal we
adopt the latest time-delay and magnification-ratio measurements from
Ref.~\cite{2023ApJ...948...93K}.  We further include 173 cluster member
galaxies, using their observed positions, ellipticities, position
angles, and luminosities relative to the brightest cluster galaxy in
the construction of the galaxy-scale mass component.

\subsection{Mass modeling}

We construct parametric strong-lensing models with the public software
{\sc glafic} \cite{2010PASJ...62.1017O,2021PASP..133g4504O}.  The
software calculates image positions, magnifications, and time delays
for a specified lens potential and optimizes the model parameters by
minimizing $\chi^2$.  To reduce the computational cost, the positional
part of $\chi^2$ is evaluated in the source plane with the full
magnification tensor taken into account.  The image positions, time
delays, and magnification or magnitude ratios described above are
included simultaneously in the fit.

\subsubsection{Mass components}

The smooth cluster-scale dark matter distribution is described by an
cored elliptical Navarro--Frenk--White profile, denoted by
\texttt{acnfw} in {\sc glafic}.  Its three-dimensional density profile
is given by
\begin{equation}
  \rho(r)=\frac{\rho_{\mathrm{s}}}{(r_{\mathrm{c}}/r_{\mathrm{s}}+r/r_{\mathrm{s}})(1+r/r_{\mathrm{s}})^2},
\end{equation}
where $\rho_{\mathrm{s}}$ is the characteristic density,
$r_{\mathrm{s}}$ is the scale radius, and $r_{\mathrm{c}}$ is the core
radius. The more detailed description and the numerical implementation 
of this density profile are given in Appendix~\ref{app:cnfw}.  
In addition to the halo mass, centroid, ellipticity, position angle, 
and concentration, the model contains the dimensionless core-size parameter
\begin{equation}
 \beta=\frac{r_{\rm c}}{r_{\rm s}}.
\end{equation}
The standard NFW profile is recovered as $\beta\rightarrow0$.  Allowing
$\beta$ to vary therefore provides a direct test of how a finite
central dark matter core affects the inferred time-delay distance.
We note that we adopt the cored NFW profile only for the main central halo,
and when multiple halos are included in the analysis we retain the standard 
NFW halo for the other halo components (see also below).

Cluster member galaxies are modeled as scaled pseudo-Jaffe ellipsoids
\cite{1983MNRAS.202..995J}, denoted by \texttt{gals}.  Their observed centroids, ellipticities, and
position angles are fixed to those measured from their light profiles 
in the imaging data, while their velocity dispersions and truncation 
radii are tied to their luminosities through
\begin{align}
 \frac{\sigma}{\sigma_*} =
 \left(\frac{L}{L_*}\right)^{1/4},\\
 \frac{r_{\rm trun}}{r_{{\rm trun},*}} =
 \left(\frac{L}{L_*}\right)^\eta ,
\end{align}
where $\sigma_*$ and $r_{{\rm trun},*}$ are the normalization parameters
for a reference galaxy of luminosity $L_*$, and $\eta$ parametrizes 
the luminosity dependence of the truncation radius.  These three
quantities are optimized together with the parameters of the
cluster-scale halos.  We also include an external shear
(\texttt{pert}) and third- and fourth-order multipole perturbations
(\texttt{mpole}) to account for angular structure not captured by the
elliptical halos and galaxy components.

We examine two treatments of the BCG.  In the first, the BCG is
included in the \texttt{gals} component and follows the same scaling
relations as the other cluster members.  In the second, it is removed
from the scaled galaxy population and modeled separately with an
elliptical Hernquist profile (\texttt{ahern}) 
\cite{1990ApJ...356..359H}.  The corresponding spherical density 
profile is
\begin{equation}
 \rho_{\rm BCG}(r)=
 \frac{M_{\rm BCG}}{2\pi}
 \frac{1}{(r/r_{\rm b})(1+r/r_{\rm b})^{3}},
\end{equation}
where $M_{\rm BCG}$ is the total mass and $r_{\rm b}$ is the scale
radius.  Both $M_{\rm BCG}$ and $r_{\rm b}$ are optimized rather 
than fixed to the observed stellar light distribution.  We use the 
fast elliptical implementation of this profile in {\sc glafic}
\cite{2021PASP..133g4504O}.

For the separate-BCG models, we consider three assumptions about
$M_{\rm BCG}$, no stellar-mass prior, a Gaussian prior on the stellar 
mass based on a stellar-population estimate with a Chabrier initial 
mass function (IMF) \cite{2003PASP..115..763C}, and a Gaussian prior 
on the stellar mass based on the corresponding estimate with a
Salpeter IMF \cite{1955ApJ...121..161S}.  Comparing these cases 
allows us to examine how the assumed BCG stellar mass affects 
constraints on the dark matter core and the inferred time-delay distance.

\subsubsection{Models of SDSS J1004+4112}

The models of SDSS J1004+4112 follow models m113 and m123 of
Ref.~\cite{2023PhRvD.108h3532L}, except that the elliptical 
NFW component \texttt{anfw} is replaced by the cored 
elliptical NFW profile \texttt{acnfw}. 
The model without a separate BCG is therefore 
\texttt{acnfw+gals+pert+mpole}(m=3,4),
whereas the model with a separate BCG is
 \texttt{acnfw+ahern+gals+pert+mpole}(m=3,4).
For the latter model, we perform fits without a BCG mass prior 
and with the Chabrier- and Salpeter-based Gaussian priors defined 
above. Based on our estimate of the stellar mass of the BCG 
described in Appendix~\ref{app:bcg_mass_1004}, we adopt 
$M_{\mathrm{BCG}}^{\mathrm{Chabrier}} = (2.27\pm1.31)\times10^{11}\,M_{\odot}/h$ and
  $M_{\mathrm{BCG}}^{\mathrm{Salpeter}}= (3.99\pm2.30)\times10^{11}\,M_{\odot}/h$ 
  for the Chabrier and Salpeter IMF cases, respectively.

\subsubsection{Models of MACS J1149.5+2223}

For MACS J1149.5+2223, we adopt the structure of model m2 in Ref.~\cite{2025PhRvD.112l3526L}.  We replace only the main NFW halo
with an \texttt{acnfw} halo; the other three cluster-scale halos retain their NFW profiles.  The external shear and the $m=3$ and $m=4$
multipole perturbations are also retained.  As in that model, the centroid and the concentration of the fourth cluster-scale halo are held fixed, and the source redshifts of the nine background systems that lack spectroscopic confirmation are optimized as free parameters.
The model without a separate BCG can thus be summarized as
 \texttt{acnfw+anfw3+gals+pert+mpole}(m=3,4).
We also construct the corresponding model in which the BCG is removed from \texttt{gals} and represented by a separate Hernquist component,
 \texttt{acnfw+anfw3+ahern+gals+pert+mpole}(m=3,4).
As in SDSS J1004+4112, the separate-BCG model is optimized with no BCG mass prior and with the Chabrier- and Salpeter-based Gaussian priors.
Based on our estimate of the stellar mass of the BCG described 
in Appendix~\ref{app:bcg_mass_1149}, we adopt 
$M_{\mathrm{BCG}}^{\mathrm{Chabrier}}
  = (3.146\pm0.815)\times10^{11}\,M_{\odot}/h$ and
  $M_{\mathrm{BCG}}^{\mathrm{Salpeter}}
  = (4.856\pm0.830)\times10^{11}\,M_{\odot}/h$ 
  for the Chabrier and Salpeter IMF cases, respectively.

\section{Results}\label{sec:results}

After constructing the lens models described in Sec.~\ref{sec:data}, we
sample the posterior with the Markov chain Monte Carlo (MCMC) 
implementation in {\sc glafic}.
The background cosmology is flat $\Lambda$CDM with a fixed matter
density $\Omega_{\rm m}=0.3$, the value used in the reference models, 
while $h$ is a free parameter.  
For each model we combine independent chains after discarding a burn-in
and thinning.  We report the median of the marginalized posterior and
the equal-tailed $68\%$ credible interval.

\subsection{SDSS J1004+4112}

\begin{figure*}[!t]
\centering
\subfigure[Scaled BCG]{%
\includegraphics[width=0.48\textwidth]{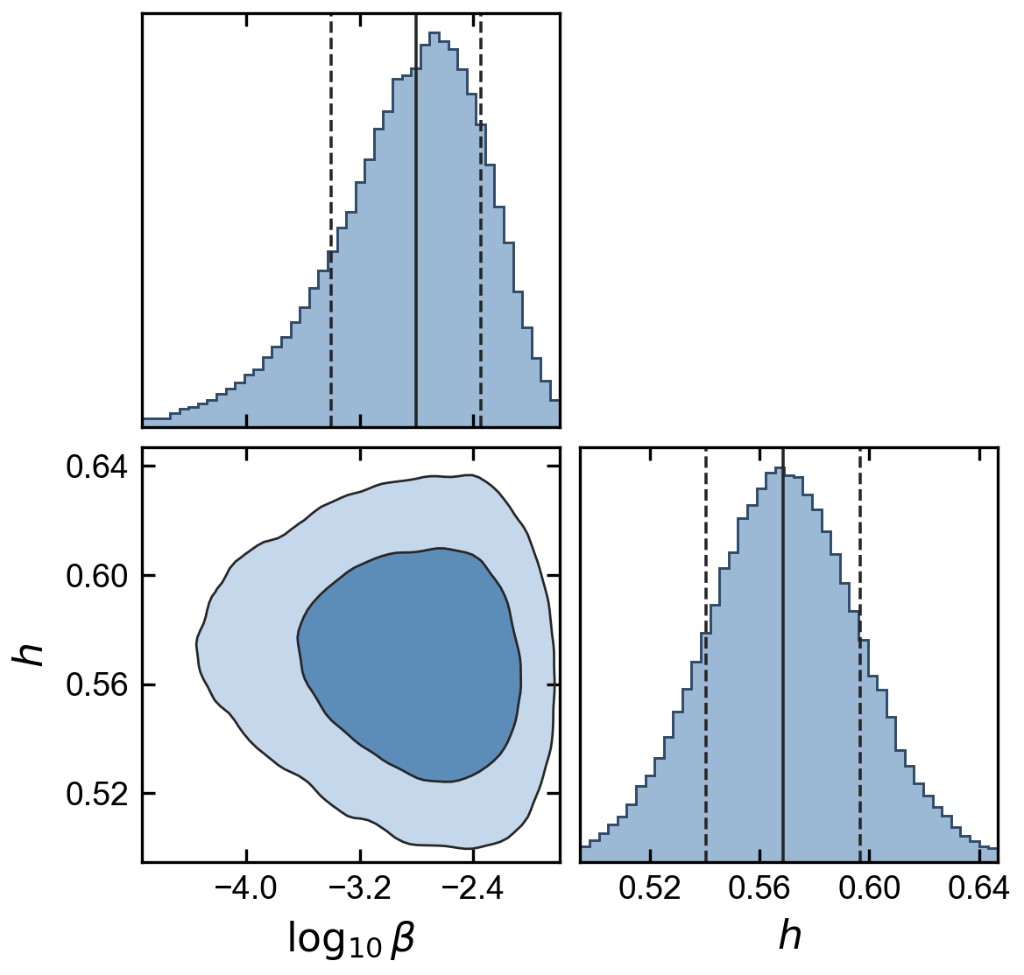}%
\label{fig:sdss1004_scaled}}
\hfill
\subfigure[No prior]{%
\includegraphics[width=0.48\textwidth]{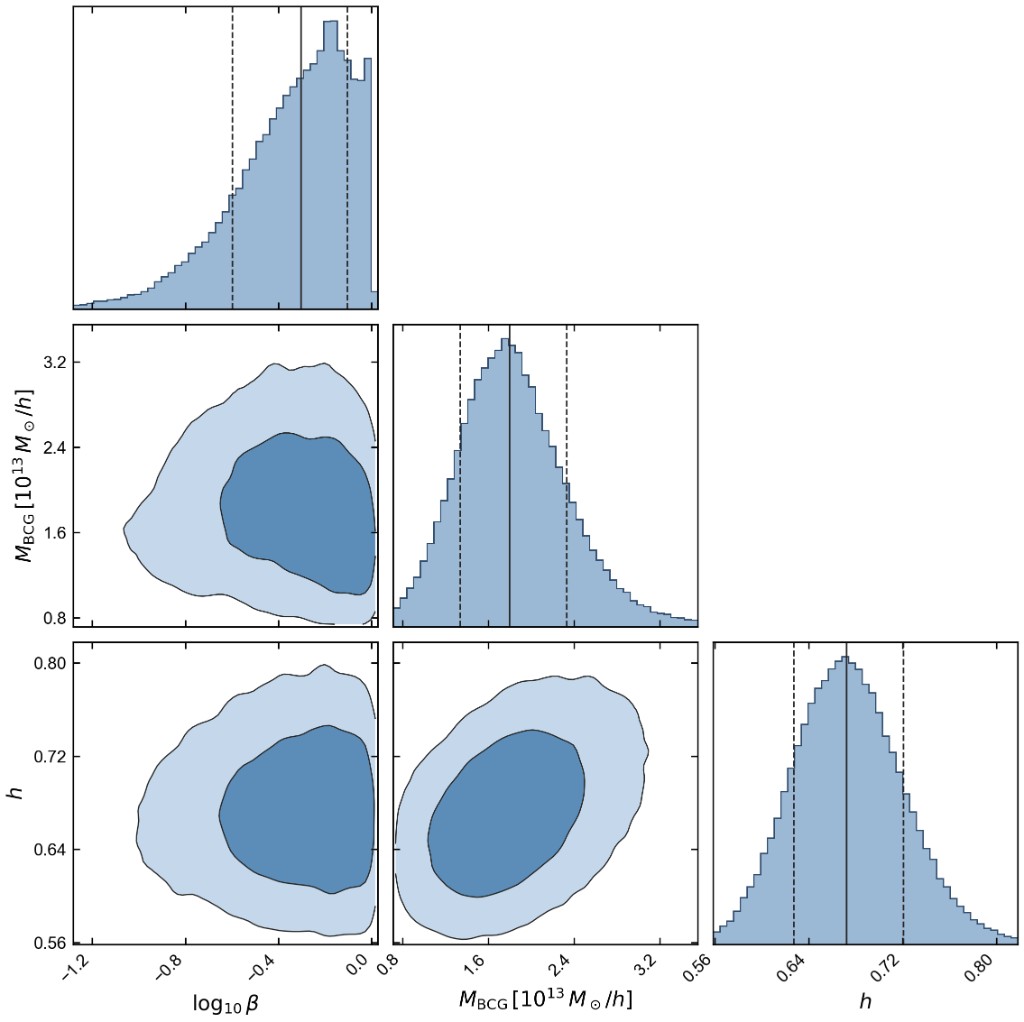}%
\label{fig:sdss1004_noprior}}
\\[0.5em]
\subfigure[Chabrier]{%
\includegraphics[width=0.48\textwidth]{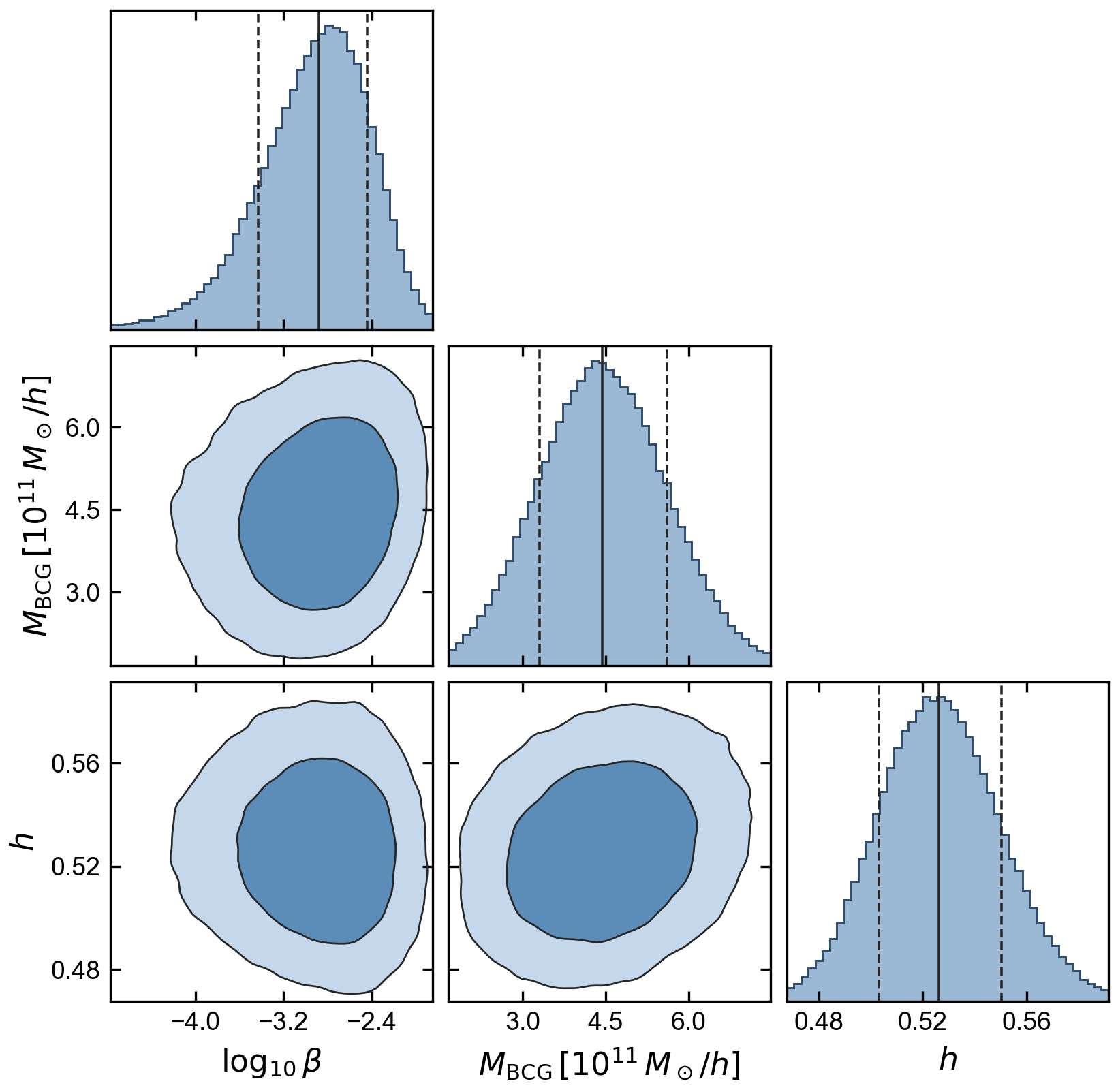}%
\label{fig:sdss1004_chabrier}}
\hfill
\subfigure[Salpeter]{%
\includegraphics[width=0.48\textwidth]{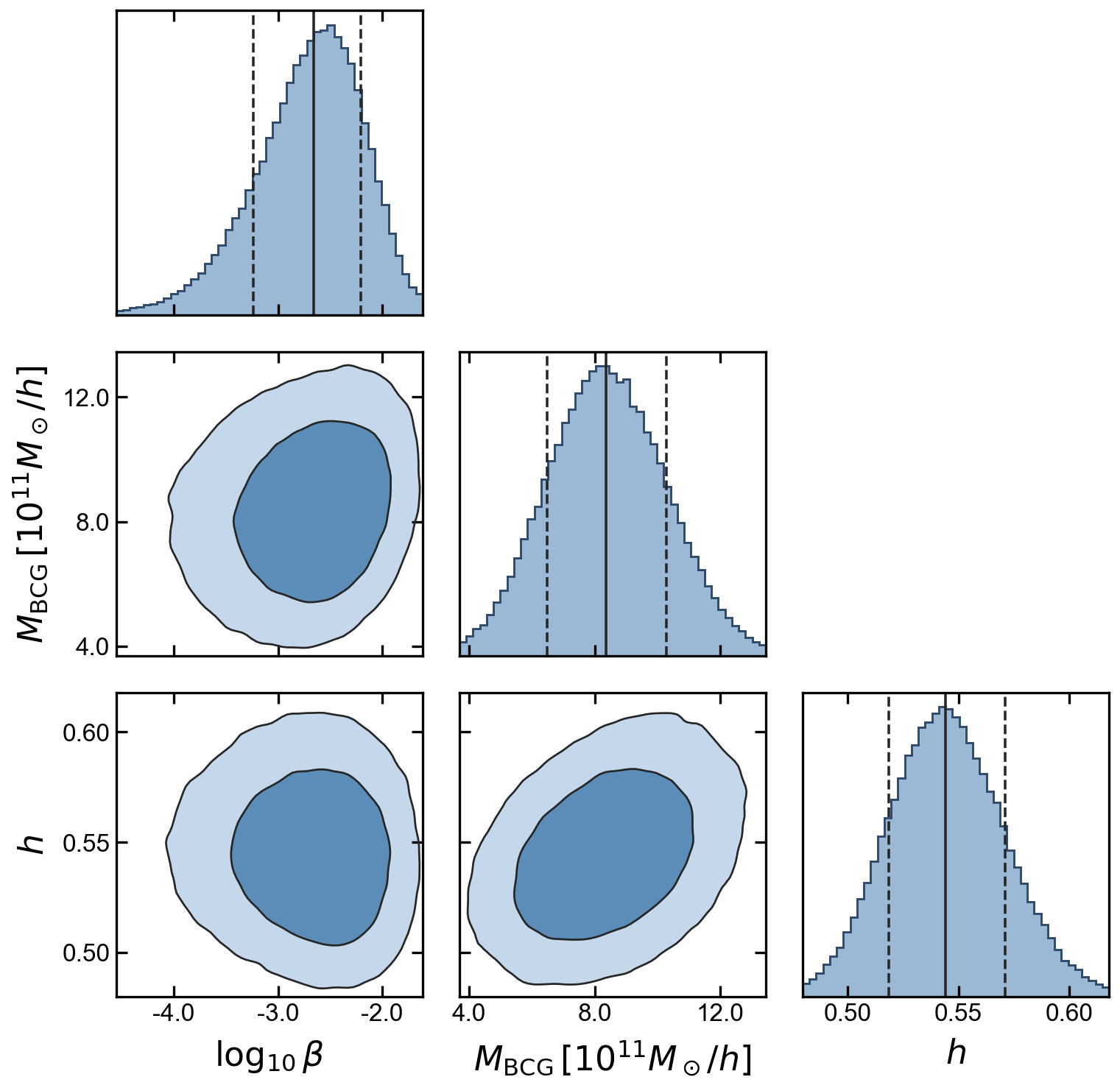}%
\label{fig:sdss1004_salpeter}}
\caption{Posterior constraints for SDSS J1004+4112 in the four different 
BCG treatments.  Panels show the joint and marginalized posteriors of
the core-size parameter $\log_{10}\beta$ and the dimensionless Hubble 
constant $h$ for the scaled-BCG model (a), and of $\log_{10}\beta$, 
the stellar mass of the BCG $M_{\rm BCG}$, and $h$ for 
the separate-BCG models with no stellar-mass prior (b), 
a Chabrier prior (c), and a Salpeter prior (d).  Contours correspond
to the $68.3\%$ and $95.5\%$ credible regions. }
\label{fig:sdss1004}
\end{figure*}

We first consider the model in which the BCG is included
in the scaled member-galaxy population.
The combined chains contain about $1.0\times10^{6}$ samples after 
thinning. The minimum $\chi^2$ of the sampled models is $40.95$.
The ``Scaled BCG'' column of Table~\ref{tab:sdss1004} summarizes 
the marginalized posteriors. The main halo has mass
$M=(5.57^{+0.51}_{-0.52})\times10^{14}\,M_\odot/h$ and concentration
$c=4.19^{+0.31}_{-0.27}$.  The core-size parameter is piled
up at small values,
$\beta=0.00159^{+0.00297}_{-0.00119}$,
or $\log_{10}\beta=-2.8^{+0.46}_{-0.60}$.   The external shear is small,
$\gamma=0.0070^{+0.0078}_{-0.0049}$, and its position angle is therefore
essentially unconstrained.
The Hubble parameter is constrained to $h=0.569\pm0.028$,
corresponding to $H_0=56.9\pm2.8\Mpc$.
Figure~\ref{fig:sdss1004}(a) shows the joint posterior of
$\log_{10}\beta$ and $h$.  The correlation between the two parameters
is weak, indicating that the low value of
$H_0$ in this model is not driven by a strong degeneracy with the core
size within the range allowed by the data.

\begin{table*}[t]
\caption{Marginalized posterior constraints for SDSS J1004+4112.  Columns
correspond to (1)~the model without a separate BCG (BCG included in
\texttt{gals}); (2)~the separate-BCG model with no stellar-mass prior; (3)~the same
separate-BCG model with a Chabrier IMF prior; and (4)~with a Salpeter
IMF prior.  We quote the median and the equal-tailed $68\%$ credible
interval.  Angles are measured East of North. At the bottom we show $\chi^2$ of
the best-fitting model and the number of degree of freedom.}
\label{tab:sdss1004}
\begin{ruledtabular}
\begin{tabular}{llcccc}
Parameter & Unit &
Scaled BCG &
No prior &
Chabrier &
Salpeter \\
\hline
$M$ & $10^{14}\,M_\odot/h$ &
$5.57^{+0.51}_{-0.52}$ &
$9.04^{+0.70}_{-1.16}$ &
$6.00\pm0.41$ &
$6.23^{+0.48}_{-0.46}$ \\
$x$ & arcsec &
$0.110^{+0.058}_{-0.056}$ &
$-0.91^{+0.36}_{-0.47}$ &
$0.050^{+0.063}_{-0.063}$ &
$-0.014^{+0.074}_{-0.076}$ \\
$y$ & arcsec &
$0.204^{+0.095}_{-0.102}$ &
$0.68^{+0.72}_{-0.49}$ &
$0.039^{+0.087}_{-0.083}$ &
$0.051^{+0.094}_{-0.089}$ \\
$e$ & --- &
$0.159^{+0.024}_{-0.020}$ &
$0.125^{+0.073}_{-0.067}$ &
$0.158^{+0.019}_{-0.019}$ &
$0.164^{+0.026}_{-0.020}$ \\
$\theta_e$ & deg &
$-22.4^{+2.6}_{-2.1}$ &
$-32^{+21}_{-23}$ &
$-21.4^{+3.2}_{-2.5}$ &
$-19.6^{+4.9}_{-4.2}$ \\
$c$ & --- &
$4.19^{+0.31}_{-0.27}$ &
$6.75^{+1.83}_{-1.72}$ &
$4.54^{+0.25}_{-0.22}$ &
$4.36^{+0.31}_{-0.26}$ \\
$\beta$ & --- &
$0.00159^{+0.00297}_{-0.00119}$ &
$0.498^{+0.290}_{-0.245}$ &
$0.00133^{+0.00232}_{-0.00096}$ &
$0.00218^{+0.00403}_{-0.00160}$ \\
$\sigma_*$ &$\kms$ &
$225^{+18}_{-21}$ &
$384^{+11}_{-22}$ &
$253^{+38}_{-39}$ &
$290^{+38}_{-40}$ \\
$r_{{\rm trun},*}$ & arcsec &
$60^{+25}_{-31}$ &
$24^{+10}_{-8}$ &
$32^{+31}_{-20}$ &
$26^{+26}_{-16}$ \\
$\eta$ & --- &
$1.31^{+0.13}_{-0.26}$ &
$1.42^{+0.06}_{-0.12}$ &
$1.32^{+0.13}_{-0.27}$ &
$1.34^{+0.12}_{-0.24}$ \\
$M_{\rm BCG}$ & $10^{11}\,M_\odot/h$ &
--- &
$180^{+53}_{-46}$ &
$4.44^{+1.16}_{-1.14}$ &
$8.36^{+1.93}_{-1.88}$ \\
$e_{\rm BCG}$ & --- &
--- &
$0.480^{+0.054}_{-0.053}$ &
$0.74^{+0.12}_{-0.25}$ &
$0.78^{+0.08}_{-0.12}$ \\
$\theta_{e,{\rm BCG}}$ & deg &
--- &
$-18.5^{+2.9}_{-1.9}$ &
$-18^{+12}_{-11}$ &
$-20.8^{+7.8}_{-8.3}$ \\
$r_{\rm b}$ & arcsec &
--- &
$8.18^{+1.46}_{-1.40}$ &
$0.92^{+0.29}_{-0.24}$ &
$1.27^{+0.34}_{-0.28}$ \\
$\gamma$ & --- &
$0.0070^{+0.0078}_{-0.0049}$ &
$0.021^{+0.019}_{-0.014}$ &
$0.0071^{+0.0070}_{-0.0049}$ &
$0.0104^{+0.0092}_{-0.0071}$ \\
$\theta_\gamma$ & deg &
$-18^{+94}_{-48}$ &
$-4^{+36}_{-37}$ &
$-28^{+99}_{-44}$ &
$-47^{+118}_{-31}$ \\
$\epsilon_3$ & --- &
$0.0186^{+0.0025}_{-0.0024}$ &
$0.0229^{+0.0028}_{-0.0030}$ &
$0.0193^{+0.0025}_{-0.0026}$ &
$0.0199^{+0.0027}_{-0.0030}$ \\
$\theta_3$ & deg &
$2.7^{+3.7}_{-3.0}$ &
$-6.3^{+2.4}_{-2.2}$ &
$-0.5^{+2.6}_{-2.3}$ &
$-1.4^{+2.6}_{-2.3}$ \\
$\epsilon_4$ & --- &
$0.0070^{+0.0022}_{-0.0017}$ &
$0.0114^{+0.0033}_{-0.0032}$ &
$0.0075^{+0.0021}_{-0.0018}$ &
$0.0077^{+0.0024}_{-0.0022}$ \\
$\theta_4$ & deg &
$51.6^{+7.2}_{-4.1}$ &
$42.7^{+2.8}_{-2.2}$ &
$139.7^{+4.6}_{-2.9}$ &
$47.6^{+4.3}_{-2.8}$ \\
$h$ & --- &
$0.569\pm0.028$ &
$0.672^{+0.048}_{-0.045}$ &
$0.526^{+0.024}_{-0.023}$ &
$0.544^{+0.027}_{-0.025}$ \\
$\chi^2/\mathrm{dof}$ & --- & $40.95/38$ & $23.27/34$ & $56.25/34$ & $47.37/34$ \\
\end{tabular}
\end{ruledtabular}
\end{table*}

\begin{figure*}[!t]
\centering
\subfigure[Core-size parameter]{%
\includegraphics[width=0.48\textwidth]{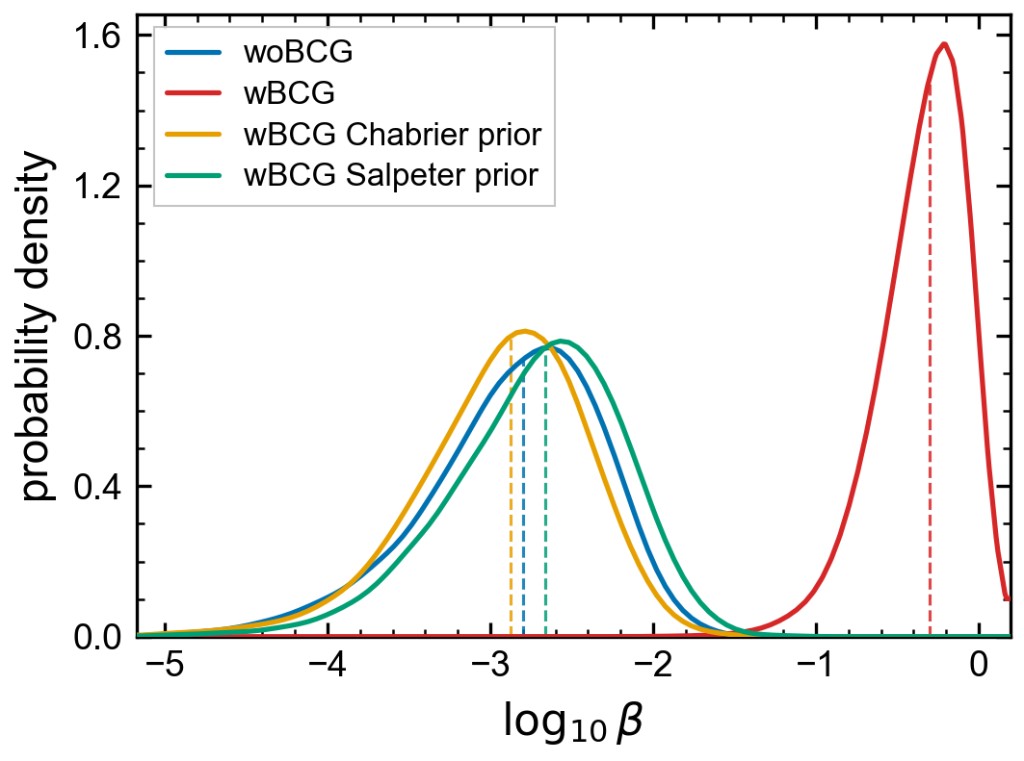}%
\label{fig:sdss1004_beta_compare}}
\hfill
\subfigure[Hubble parameter]{%
\includegraphics[width=0.48\textwidth]{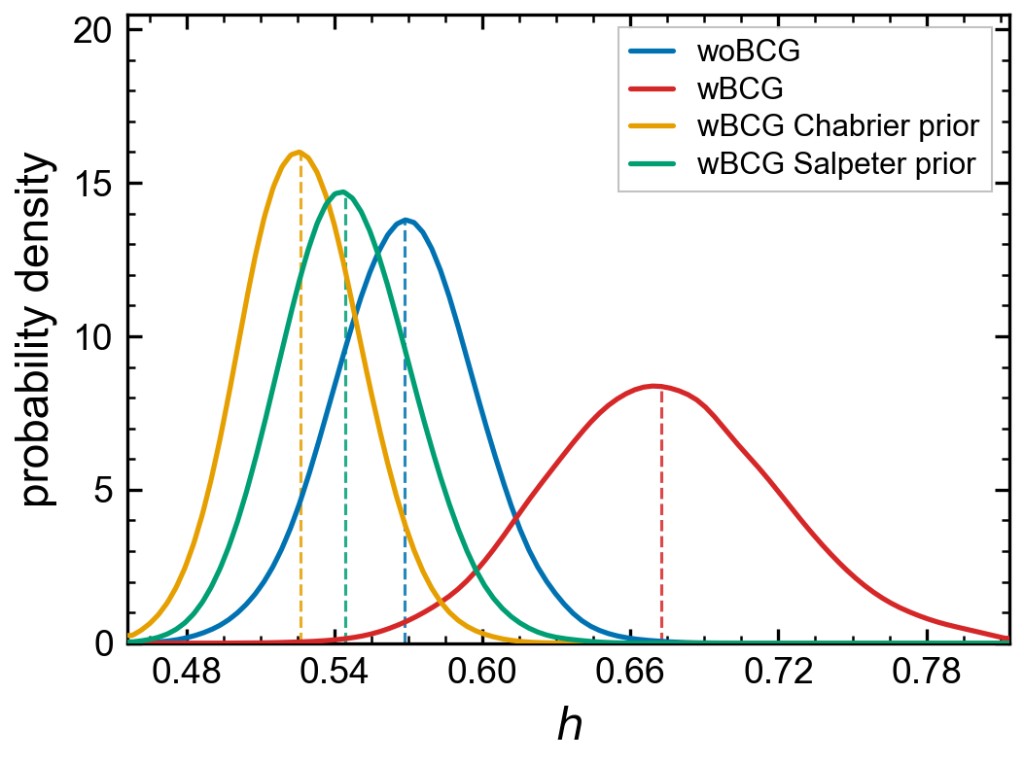}%
\label{fig:sdss1004_h_compare}}
\caption{Comparison of the one-dimensional marginalized posterior
densities of the core-size parameter $\log_{10}\beta$ (a) and 
the dimensionless Hubble constant $h$ (b) for the four
SDSS J1004+4112 models.  The dashed vertical lines mark the posterior
medians.  In the legends, ``woBCG'' denotes the model in which the BCG
is included in the scaled member-galaxy population, rather than a
model without a BCG mass component.  Since each curve is normalized
separately, its height does not measure the relative evidence of the 
corresponding model.}
\label{fig:sdss1004_compare}
\end{figure*}

We next consider the model in which the BCG is
removed from the scaled galaxy population and represented by a
separate Hernquist component, without imposing a stellar-mass prior on
$M_{\rm BCG}$.  The combined chains contain about $9.4\times10^{5}$
samples after thinning. Without a mass prior, a subset of samples has
the main-halo centroid displaced by more than $10''$ from the cluster
center, indicating that the halo is no longer sitting on the BCG, 
and the fit is using the extra central freedom to place mass well 
away from the light. We therefore report the posterior restricted to the
on-cluster subset $|x|<10''$ and $|y|<10''$, which retains about
$8.0\times10^{5}$ samples ($\simeq85\%$ of the chain). The global 
minimum $\chi^2$ among the MCMC samples is $23.27$, lower than the 
$40.95$ of the scaled-BCG model, as expected from the four 
additional BCG parameters, $M_{\rm BCG}$, $e_{\rm BCG}$, 
$\theta_{e,{\rm BCG}}$, and $r_{\rm b}$.

The marginalized posteriors are listed in the ``No prior'' column of
Table~\ref{tab:sdss1004}.  The main halo mass and concentration are
$M=(9.04^{+0.70}_{-1.16})\times10^{14}\,M_\odot/h$ and
$c=6.75^{+1.83}_{-1.72}$.  The core-size parameter is significantly 
non-zero, $\beta=0.498^{+0.290}_{-0.245}$.
The BCG mass is
$M_{\rm BCG}=(1.80^{+0.53}_{-0.46})\times10^{13}\,M_\odot/h,$
about two orders of magnitude above the Chabrier and Salpeter
stellar-population estimates used as priors below.  The scale radius
of the Hernquist component is
$r_{\rm b}=8.18^{+1.46}_{-1.40}$~arcsec,
several times a typical BCG effective radius, and the remaining
member-galaxy normalization rises to
$\sigma_*=384^{+11}_{-22}\,\kms$.  Both values show that, once
$M_{\rm BCG}$ is free, the fit absorbs central mass into an oversized
stellar component and into the scaled galaxy population rather than
into a physically motivated BCG. The dimensionless Hubble parameter is
$h=0.672^{+0.048}_{-0.045}$.
Compared with the model without a separate BCG, both $\beta$ and $H_0$
are substantially larger.  Figure~\ref{fig:sdss1004}(b)
shows the joint posterior of $\log_{10}\beta$, $M_{\rm BCG}$, and $h$.
There is a clear positive correlation between $M_{\rm BCG}$ and $h$,
while $\beta$ is only weakly correlated with either quantity.

We then impose a Gaussian prior based on a Chabrier IMF stellar-mass
estimate on $M_{\rm BCG}$ of the same separate-BCG model,
$M_{\rm BCG}^{\rm Chabrier}=(2.27\pm1.31)\times10^{11}\,M_\odot/h$.
The combined chains contain about $9.7\times10^{5}$ samples after
thinning.   The minimum $\chi^2$ of the sampled models is $56.25$.  
The marginalized posteriors are listed in the ``Chabrier'' column of
Table~\ref{tab:sdss1004}.  The main halo parameters remain broadly
consistent with those of the no-separate-BCG model, with
$M=(6.00\pm0.41)\times10^{14}\,M_\odot/h$ and
$c=4.54^{+0.25}_{-0.22}$.  The core-size parameter is again piled up at
small values, $\beta=(1.33^{+2.32}_{-0.96})\times10^{-3}$.
 The BCG mass is constrained to
$M_{\rm BCG}=(4.44^{+1.16}_{-1.14})\times10^{11}\,M_\odot/h$,
which is $1.2\sigma$ above the Chabrier prior mean, indicating that the
lensing constraints prefer a more massive central galaxy than the
stellar-population estimate alone.  The dimensionless Hubble parameter is
$h=0.526^{+0.024}_{-0.023}$. Compared with to the model without 
a BCG mass prior, both $\beta$ and $H_0$
are substantially smaller and closer to the values obtained when the
BCG is not modeled separately.  Figure~\ref{fig:sdss1004}(c)
shows the joint posterior of $\log_{10}\beta$, $M_{\rm BCG}$, and $h$.
Within the range allowed by the data, neither $\beta$ nor
$M_{\rm BCG}$ exhibits a strong correlation with $h$.

We repeat the separate-BCG analysis with a Gaussian prior based on a
Salpeter IMF stellar-mass estimate,
$M_{\rm BCG}^{\rm Salpter}=(3.99\pm2.30)\times10^{11}\,M_\odot/h$.
The combined chains contain about $9.7\times10^{5}$ samples after
thinning. The minimum $\chi^2$ of the sampled models is $47.37$.
The marginalized posteriors are listed in the ``Salpeter'' column of
Table~\ref{tab:sdss1004}.  The main halo mass and concentration are
$M=(6.23^{+0.48}_{-0.46})\times10^{14}\,M_\odot/h$ and
$c=4.36^{+0.31}_{-0.26}$.  The core-size parameter is
$\beta=(2.18^{+4.03}_{-1.60})\times10^{-3}$.  
The BCG mass is constrained to $M_{\rm BCG}=(8.36^{+1.93}_{-1.88})\times10^{11}\,M_\odot/h$,
again higher than the prior mean.  The dimensionless 
Hubble parameter is $h=0.544^{+0.027}_{-0.025}$.
Compared with the Chabrier-prior model, the Salpeter prior allows a
larger BCG mass and yields a slightly larger $\beta$ and a slightly
larger $H_0$.  Both remain much smaller than in the model without a
BCG mass prior. Compared with the model without a separate BCG, $H_0$
remains lower, while $\beta$ is of the same order and still consistent
with a nearly cuspy profile.  Figure~\ref{fig:sdss1004}(d)
shows the joint posterior of $\log_{10}\beta$, $M_{\rm BCG}$, and $h$.
A mild positive correlation between $M_{\rm BCG}$ and $h$ is visible.

Figure~\ref{fig:sdss1004_compare} provides a direct comparison of the
marginalized constraints from the four different BCG treatments.  
When the BCG mass is allowed to vary without a stellar-mass prior, the
$\log_{10}\beta$ posterior shifts from the nearly cuspy regime occupied
by the other three models to values near $-0.3$, corresponding to
$\beta$ of order unity and $r_{\rm c}\simeq100\,h^{-1}\,\mathrm{kpc}$.
The scaled-BCG, Chabrier-prior, and
Salpeter-prior posteriors overlap substantially such that all three 
pile up near $\beta\sim10^{-3}$ ($r_{\rm c}\lesssim1\,h^{-1}\,\mathrm{kpc}$)
and are limited from below by the prior volume rather than by the
lensing data.

The same change in the central mass decomposition
also shifts the $h$ posterior.  The no-prior model gives the highest
median, $H_0=67.2^{+4.8}_{-4.5}\Mpc$, consistent with the 16-model NFW
analysis of the same cluster, $H_0=67.5^{+14.5}_{-8.9}\Mpc$
\cite{2023PhRvD.108h3532L}.  The Chabrier and Salpeter priors give
$H_0=52.6^{+2.4}_{-2.3}\Mpc$ and
$54.4^{+2.7}_{-2.5}\Mpc$, a downward shift of $\simeq2.5\sigma$
relative to the unconstrained separate-BCG model, to values below both
the Planck and the Cepheid anchors.  The scaled-BCG result,
$H_0=56.9\pm2.8\Mpc$, lies between the stellar-mass-prior and no-prior
solutions.  These one-dimensional comparisons show that the shifts in
$\beta$ and $h$ occur together when the BCG prescription is changed. 
We note that they do not establish a direct degeneracy between the
two parameters, which is weak within each individual model
(see Fig.~\ref{fig:sdss1004}).

\subsection{MACS J1149.5+2223}

\begin{figure*}[!t]
\centering
\subfigure[Scaled BCG]{%
\includegraphics[width=0.48\textwidth]{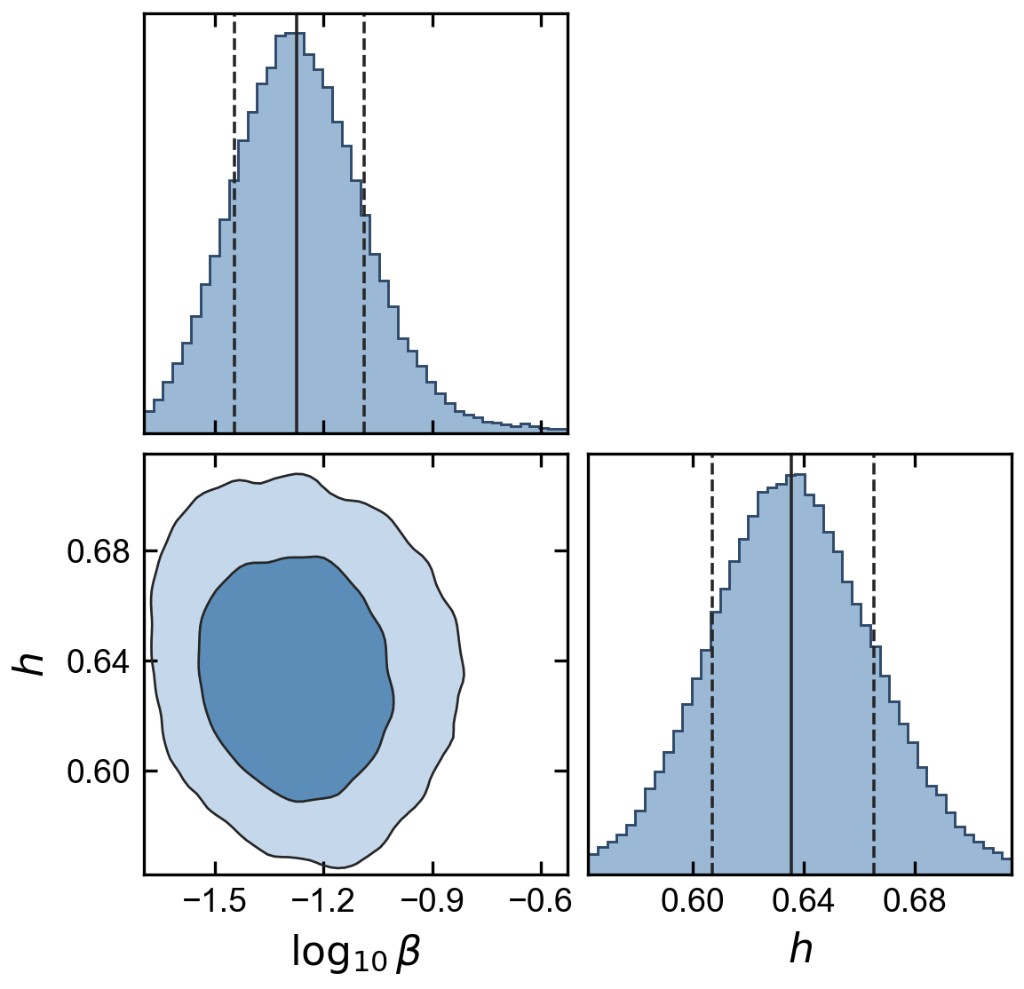}%
\label{fig:macsj1149_scaled}}
\hfill
\subfigure[No prior]{%
\includegraphics[width=0.48\textwidth]{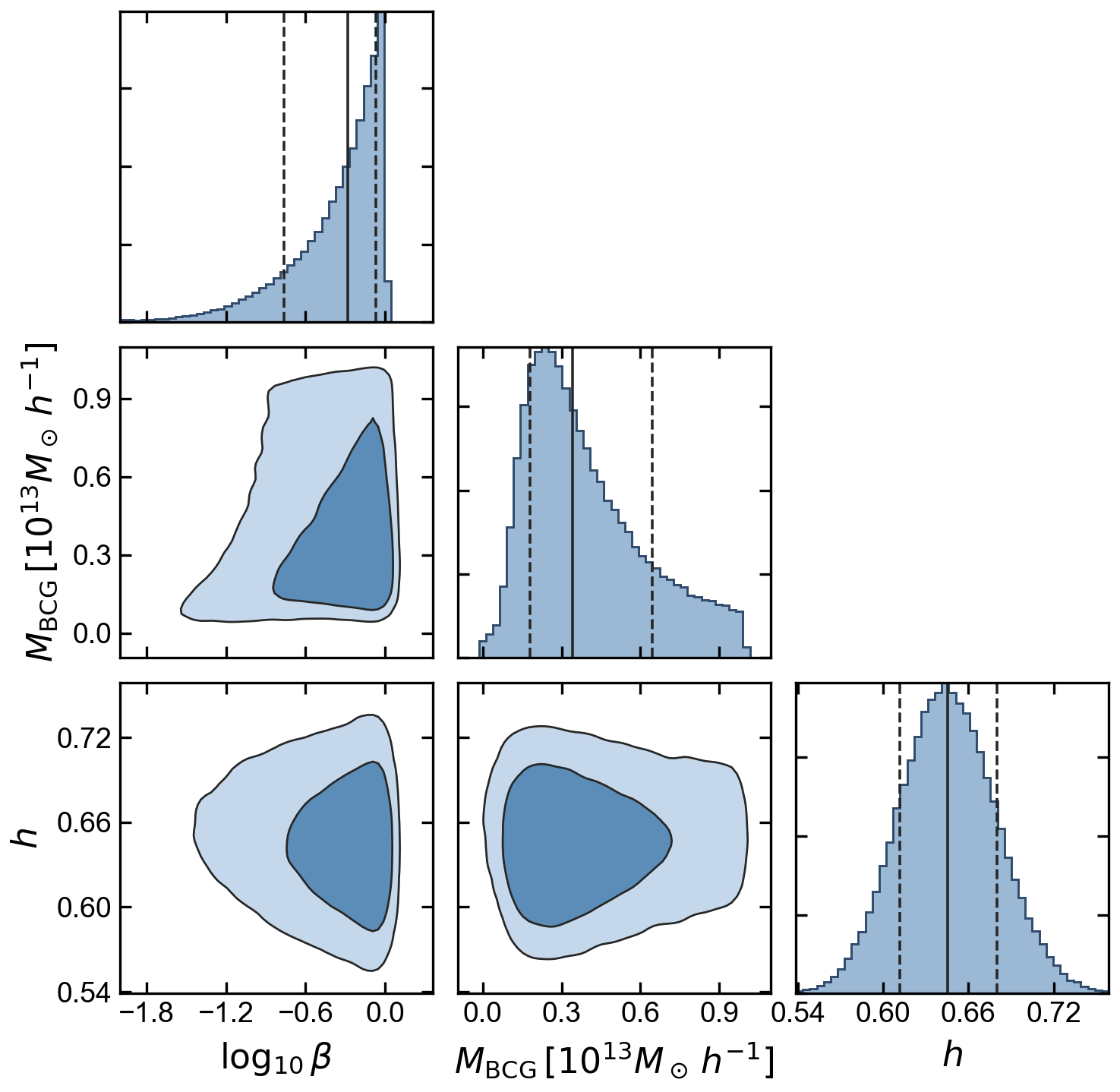}%
\label{fig:macsj1149_noprior}}
\\[0.5em]
\subfigure[Chabrier]{%
\includegraphics[width=0.48\textwidth]{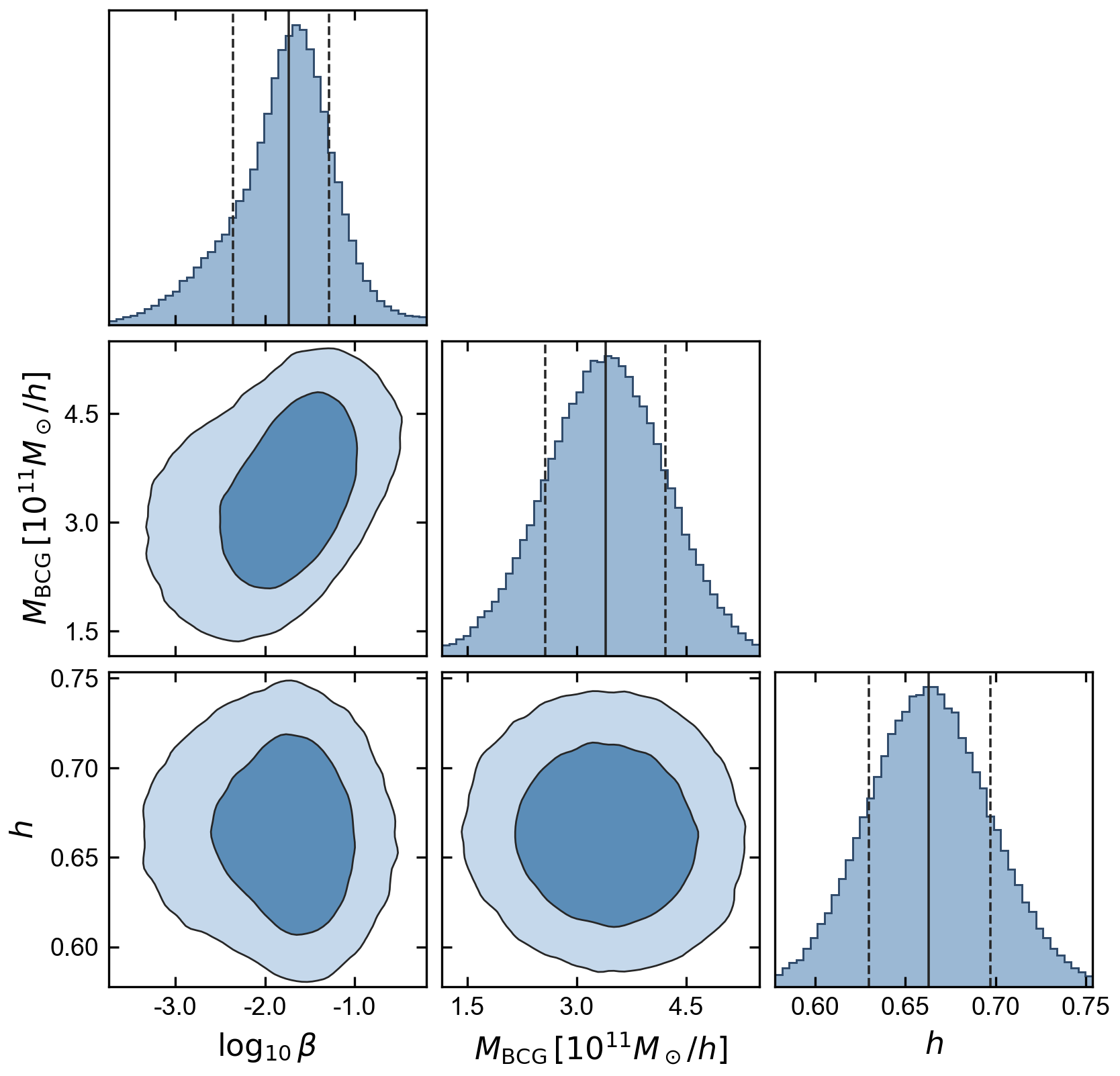}%
\label{fig:macsj1149_chabrier}}
\hfill
\subfigure[Salpeter]{%
\includegraphics[width=0.48\textwidth]{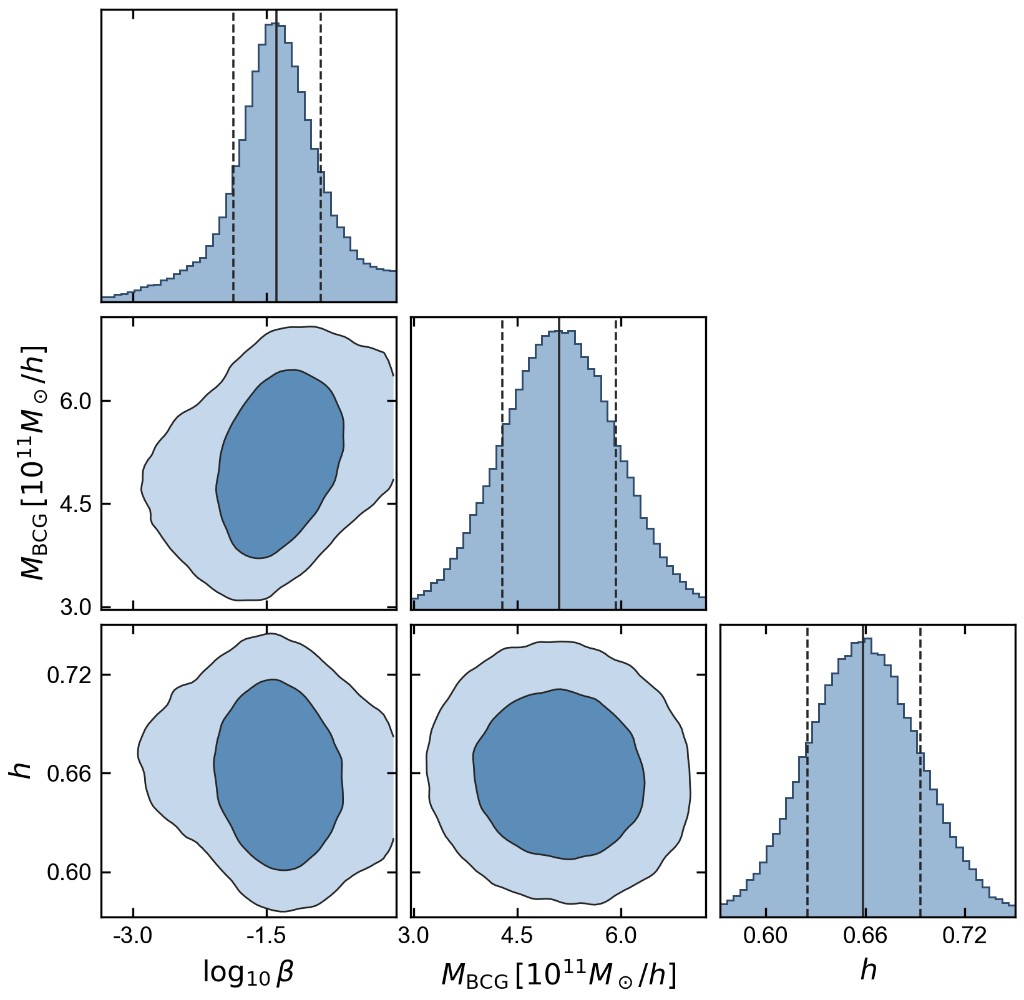}%
\label{fig:macsj1149_salpeter}}
\caption{Similar to Fig.~\ref{fig:sdss1004}, but for MACS J1149.5+2223.}
\label{fig:macsj1149}
\end{figure*}

We first consider the model in which the BCG is included in the 
scaled member-galaxy population.  The combined chains contain 
about $2.8\times10^{5}$ samples after thinning. The global minimum 
$\chi^2$ is $67.27$. The ``Scaled BCG'' column of 
Table~\ref{tab:macsj1149} summarizes the marginalized posteriors,
and Fig.~\ref{fig:macsj1149}(a) shows the joint posterior of
$\log_{10}\beta$ and $h$.  The main \texttt{acnfw} halo has mass
$M_1=(4.72^{+0.74}_{-0.91})\times10^{14}\,M_\odot/h$ and concentration
$c_1=6.17^{+0.86}_{-0.59}$.  The core-size parameter is
$\beta=0.0534^{+0.0285}_{-0.0175}$, or equivalently
$\log_{10}\beta=-1.27^{+0.19}_{-0.17}$.
Unlike the corresponding model of SDSS J1004+4112, the posterior does
not pile up at small $\beta$. It is away from zero within 
the $68\%$ interval.   The dimensionless Hubble parameter is 
constrained to $h=0.635^{+0.030}_{-0.028}$. 
Both $\beta$ and $H_0$ are substantially larger than in the
no-separate-BCG model of SDSS J1004+4112.  The correlation between
$\log_{10}\beta$ and $h$ is weak, indicating that the higher value of
$H_0$ is not driven by a tight degeneracy with the core size within
the range allowed by the data.

\begin{table*}[t]
\caption{Marginalized posterior constraints for MACS J1149.5+2223.
Columns correspond to (1)~the model without a separate BCG (BCG
included in \texttt{gals}); (2)~the separate-BCG model with no
stellar-mass prior; (3)~the same separate-BCG model with a Chabrier
IMF prior; and (4)~with a Salpeter IMF prior.  We quote the median
and the equal-tailed $68\%$ credible interval.  Subscripts $1$--$4$
label the main \texttt{acnfw} halo and the three additional NFW halos.
The centroid and concentration of halo~4 are held fixed.  Subscript
$J$ denotes the cluster member responsible for the Einstein-cross
images of SN Refsdal.  Angles are measured East of North.  Entries
marked ``---'' are not free parameters in that model. At the bottom 
we show $\chi^2$ of the best-fitting model and the number of degree of freedom.}
\label{tab:macsj1149}
\begin{ruledtabular}
\centering
{\scriptsize
\setlength{\tabcolsep}{2.5pt}
\renewcommand{\arraystretch}{1.04}
\begin{adjustbox}{max width=\textwidth}
\begin{tabular}{llcccc}
Parameter & Unit &
Scaled BCG &
No prior &
Chabrier &
Salpeter \\
\hline
$M_1$ & $10^{14}\,M_\odot/h$ &
$4.72^{+0.74}_{-0.91}$ &
$5.28^{+1.39}_{-1.21}$ &
$4.45^{+1.35}_{-1.15}$ &
$4.26^{+1.39}_{-1.12}$ \\
$x_1$ & arcsec &
$-0.194^{+0.116}_{-0.132}$ &
$-0.025^{+0.403}_{-0.366}$ &
$-0.032^{+0.177}_{-0.188}$ &
$-0.071^{+0.209}_{-0.230}$ \\
$y_1$ & arcsec &
$0.005^{+0.132}_{-0.098}$ &
$0.408^{+0.281}_{-0.261}$ &
$0.199^{+0.147}_{-0.130}$ &
$0.245^{+0.157}_{-0.146}$ \\
$e_1$ & --- &
$0.450^{+0.020}_{-0.019}$ &
$0.461^{+0.026}_{-0.027}$ &
$0.448^{+0.021}_{-0.022}$ &
$0.448^{+0.022}_{-0.023}$ \\
$\theta_{e,1}$ & deg &
$-53.3\pm1.2$ &
$-51.2^{+1.5}_{-1.9}$ &
$-52.7^{+1.5}_{-1.8}$ &
$-52.5^{+1.4}_{-1.8}$ \\
$c_1$ & --- &
$6.17^{+0.86}_{-0.59}$ &
$10.6^{+2.3}_{-3.3}$ &
$5.51^{+1.31}_{-0.74}$ &
$6.13^{+2.22}_{-1.08}$ \\
$\beta$ & --- &
$0.0534^{+0.0285}_{-0.0175}$ &
$0.523^{+0.331}_{-0.350}$ &
$0.0184^{+0.0337}_{-0.0140}$ &
$0.0410^{+0.0865}_{-0.0276}$ \\
$M_2$ & $10^{14}\,M_\odot/h$ &
$9.24^{+1.91}_{-1.67}$ &
$8.44^{+5.15}_{-3.41}$ &
$9.94^{+5.87}_{-3.80}$ &
$10.50^{+6.47}_{-4.13}$ \\
$x_2$ & arcsec &
$7.25\pm1.06$ &
$8.03^{+1.81}_{-1.51}$ &
$8.20^{+1.91}_{-1.81}$ &
$8.09^{+1.83}_{-1.75}$ \\
$y_2$ & arcsec &
$67.76^{+0.92}_{-1.07}$ &
$66.2^{+6.1}_{-5.5}$ &
$67.3^{+5.9}_{-4.8}$ &
$67.5^{+6.2}_{-5.0}$ \\
$e_2$ & --- &
$0.635^{+0.068}_{-0.059}$ &
$0.535^{+0.146}_{-0.206}$ &
$0.588^{+0.119}_{-0.152}$ &
$0.571^{+0.128}_{-0.162}$ \\
$\theta_{e,2}$ & deg &
$85.1^{+4.7}_{-4.3}$ &
$84.5^{+12.5}_{-12.9}$ &
$76.8^{+7.2}_{-7.3}$ &
$77.7^{+7.8}_{-7.9}$ \\
$c_2$ & --- &
$4.04^{+0.81}_{-0.83}$ &
$3.09^{+1.49}_{-1.1}$ &
$3.00^{+1.19}_{-0.92}$ &
$2.93^{+1.24}_{-0.93}$ \\
$M_3$ & $10^{13}\,M_\odot/h$ &
$6.22^{+8.98}_{-3.04}$ &
$3.21^{+5.90}_{-2.21}$ &
$5.57^{+8.22}_{-3.30}$ &
$5.09^{+8.25}_{-3.16}$ \\
$x_3$ & arcsec &
$-24.70^{+0.78}_{-0.71}$ &
$-24.8^{+1.12}_{-1.04}$ &
$-24.43^{+1.07}_{-0.99}$ &
$-24.56^{+1.06}_{-0.99}$ \\
$y_3$ & arcsec &
$-32.42^{+1.06}_{-1.08}$ &
$-32.7^{+1.44}_{-1.43}$ &
$-32.08^{+1.39}_{-1.26}$ &
$-32.23^{+1.35}_{-1.24}$ \\
$e_3$ & --- &
$0.622^{+0.053}_{-0.069}$ &
$0.613^{+0.089}_{-0.131}$ &
$0.666^{+0.064}_{-0.086}$ &
$0.646^{+0.074}_{-0.103}$ \\
$\theta_{e,3}$ & deg &
$-28.8^{+3.7}_{-3.0}$ &
$-26.1^{+10}_{-5.7}$ &
$-28.5^{+5.7}_{-3.9}$ &
$-28.6^{+6.6}_{-4.2}$ \\
$c_3$ & --- &
$3.13^{+1.43}_{-1.11}$ &
$3.90^{+3.82}_{-1.55}$ &
$3.17^{+1.81}_{-1.08}$ &
$3.32^{+2.17}_{-1.19}$ \\
$M_4$ & $10^{13}\,M_\odot/h$ &
$1.65^{+0.45}_{-0.37}$ &
$1.92^{+0.72}_{-0.65}$ &
$1.55^{+0.62}_{-0.54}$ &
$1.61^{+0.66}_{-0.58}$ \\
$e_4$ & --- &
$0.748^{+0.038}_{-0.082}$ &
$0.749^{+0.037}_{-0.098}$ &
$0.757^{+0.032}_{-0.089}$ &
$0.752^{+0.036}_{-0.108}$ \\
$\theta_{e,4}$ & deg &
$-33.0^{+3.0}_{-2.7}$ &
$-34.8^{+4.7}_{-3.6}$ &
$-32.7^{+4.1}_{-3.5}$ &
$-33.4^{+4.4}_{-3.7}$ \\
$\sigma_*$ & $\kms$ &
$270^{+13}_{-12}$ &
$290^{+23}_{-18}$ &
$291^{+26}_{-20}$ &
$289^{+24}_{-19}$ \\
$r_{\rm trun,*}$ & arcsec &
$5.70^{+1.73}_{-1.44}$ &
$7.68^{+8.81}_{-3.72}$ &
$5.77^{+5.75}_{-2.60}$ &
$6.24^{+6.34}_{-2.84}$ \\
$\eta$ & --- &
$0.48^{+0.12}_{-0.12}$ &
$0.64^{+0.28}_{-0.22}$ &
$0.61^{+0.26}_{-0.21}$ &
$0.62^{+0.27}_{-0.22}$ \\
$M_{\rm BCG}$ & $10^{11}\,M_\odot/h$ &
--- &
$34.1^{+30.3}_{-16.1}$ &
$3.40^{+0.82}_{-0.83}$ &
$5.10\pm0.82$ \\
$e_{\rm BCG}$ & --- &
--- &
$0.207^{+0.194}_{-0.142}$ &
$0.51^{+0.25}_{-0.32}$ &
$0.44^{+0.23}_{-0.27}$ \\
$\theta_{e,{\rm BCG}}$ & deg &
--- &
$-50.4^{+95}_{-87}$ &
$7^{+32}_{-40}$ &
$6^{+27}_{-36}$ \\
$r_{\rm b}$ & arcsec &
--- &
$5.94^{+2.92}_{-2.11}$ &
$1.52^{+2.17}_{-0.55}$ &
$1.71^{+0.91}_{-0.45}$ \\
$\sigma_J$ & $\kms$ &
$221^{+21}_{-13}$ &
$236^{+54}_{-24}$ &
$240^{+59}_{-25}$ &
$237^{+55}_{-24}$ \\
$e_J$ & --- &
$0.063^{+0.068}_{-0.042}$ &
$0.087^{+0.097}_{-0.059}$ &
$0.081^{+0.095}_{-0.056}$ &
$0.077^{+0.092}_{-0.053}$ \\
$\theta_{e,J}$ & deg &
$-170^{+18}_{-15}$ &
$-82^{+291}_{-205}$ &
$-89^{+288}_{-202}$ &
$-77^{+282}_{-209}$ \\
$r_{s,J}$ & arcsec &
$1.36^{+0.44}_{-0.38}$ &
$1.07^{+0.57}_{-0.49}$ &
$1.03^{+0.55}_{-0.49}$ &
$1.08^{+0.57}_{-0.50}$ \\
$\gamma$ & --- &
$0.076^{+0.012}_{-0.010}$ &
$0.0744^{+0.018}_{-0.017}$ &
$0.073^{+0.017}_{-0.016}$ &
$0.075^{+0.018}_{-0.016}$ \\
$\theta_\gamma$ & deg &
$277.2^{+5.4}_{-5.1}$ &
$280.0^{+9.1}_{-7.7}$ &
$282.8^{+9.3}_{-8.4}$ &
$283.0^{+9.2}_{-8.5}$ \\
$\epsilon_3$ & --- &
$0.0233^{+0.0022}_{-0.0023}$ &
$0.0187^{+0.0053}_{-0.0047}$ &
$0.0222^{+0.0040}_{-0.0039}$ &
$0.0215\pm0.0040$ \\
$\theta_3$ & deg &
$95.8^{+2.6}_{-3.2}$ &
$99.5^{+4.5}_{-4.6}$ &
$99.4\pm3.8$ &
$99.4^{+4.0}_{-3.9}$ \\
$\epsilon_4$ & --- &
$0.00132^{+0.00021}_{-0.00015}$ &
$0.000014^{+0.00068}_{-0.000014}$ &
$0.00004^{+0.00073}_{-0.00004}$ &
$0.00004^{+0.00075}_{-0.00004}$ \\
$\theta_4$ & deg &
$66.4^{+8.8}_{-8.2}$ &
$81^{+180}_{-250}$ &
$71^{+178}_{-201}$ &
$72^{+173}_{-200}$ \\
$h$ & --- &
$0.635^{+0.030}_{-0.028}$ &
$0.645^{+0.035}_{-0.034}$ &
$0.663^{+0.034}_{-0.033}$ &
$0.659^{+0.035}_{-0.034}$ \\
$z_{14}$ & --- &
$2.722^{+0.081}_{-0.077}$ &
$2.739^{+0.094}_{-0.081}$ &
$2.750^{+0.087}_{-0.080}$ &
$2.742^{+0.087}_{-0.079}$ \\
$z_{17}$ & --- &
$1.47^{+1.30}_{-0.97}$ &
$4.91^{+3.43}_{-3.45}$ &
$4.91^{+3.44}_{-3.47}$ &
$4.91^{+3.44}_{-3.48}$ \\
$z_{19}$ & --- &
$2.719^{+0.089}_{-0.085}$ &
$2.761^{+0.090}_{-0.086}$ &
$2.757^{+0.090}_{-0.085}$ &
$2.759^{+0.090}_{-0.085}$ \\
$z_{20}$ & --- &
$3.25^{+0.11}_{-0.10}$ &
$3.21^{+0.11}_{-0.11}$ &
$3.24^{+0.11}_{-0.11}$ &
$3.23^{+0.11}_{-0.11}$ \\
$z_{26}$ & --- &
$1.809^{+0.034}_{-0.035}$ &
$1.814^{+0.039}_{-0.038}$ &
$1.809^{+0.037}_{-0.036}$ &
$1.814^{+0.038}_{-0.037}$ \\
$z_{28}$ & --- &
$6.51^{+0.23}_{-0.32}$ &
$6.43^{+0.42}_{-0.41}$ &
$6.42^{+0.42}_{-0.40}$ &
$6.42^{+0.42}_{-0.40}$ \\
$z_{30}$ & --- &
$1.335^{+0.069}_{-0.067}$ &
$1.347^{+0.099}_{-0.074}$ &
$1.353^{+0.098}_{-0.074}$ &
$1.351^{+0.097}_{-0.073}$ \\
$z_{31}$ & --- &
$1.389^{+0.058}_{-0.059}$ &
$1.392^{+0.074}_{-0.063}$ &
$1.402^{+0.073}_{-0.063}$ &
$1.400^{+0.073}_{-0.063}$ \\
$z_{32}$ & --- &
$8.40^{+0.45}_{-0.47}$ &
$8.39^{+0.50}_{-0.50}$ &
$8.39\pm0.50$ &
$8.39\pm0.49$ \\
$\chi^2/\mathrm{dof}$ & --- & $67.27/116$ & $70.59/112$ & $71.80/112$ & $71.81/112$\\
\end{tabular}
\end{adjustbox}
}
\end{ruledtabular}
\end{table*}

\begin{figure*}[!t]
\centering
\subfigure[Core-size parameter]{%
\includegraphics[width=0.48\textwidth]{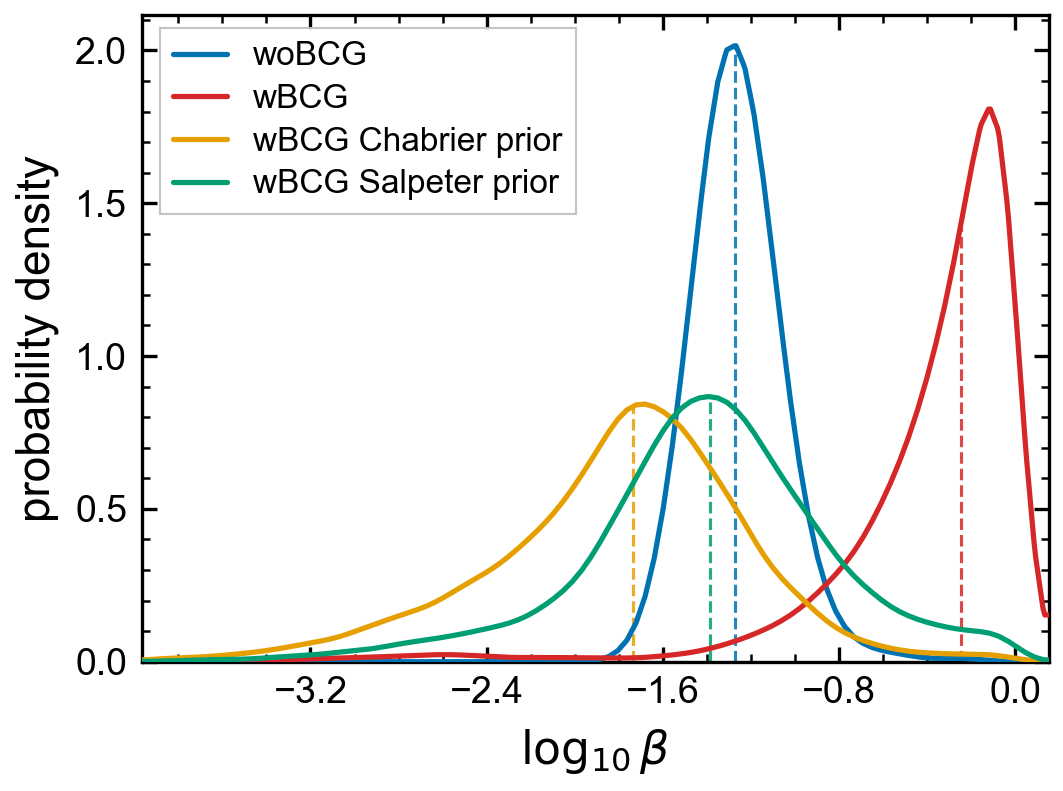}%
\label{fig:macsj1149_beta_compare}}
\hfill
\subfigure[Hubble parameter]{%
\includegraphics[width=0.48\textwidth]{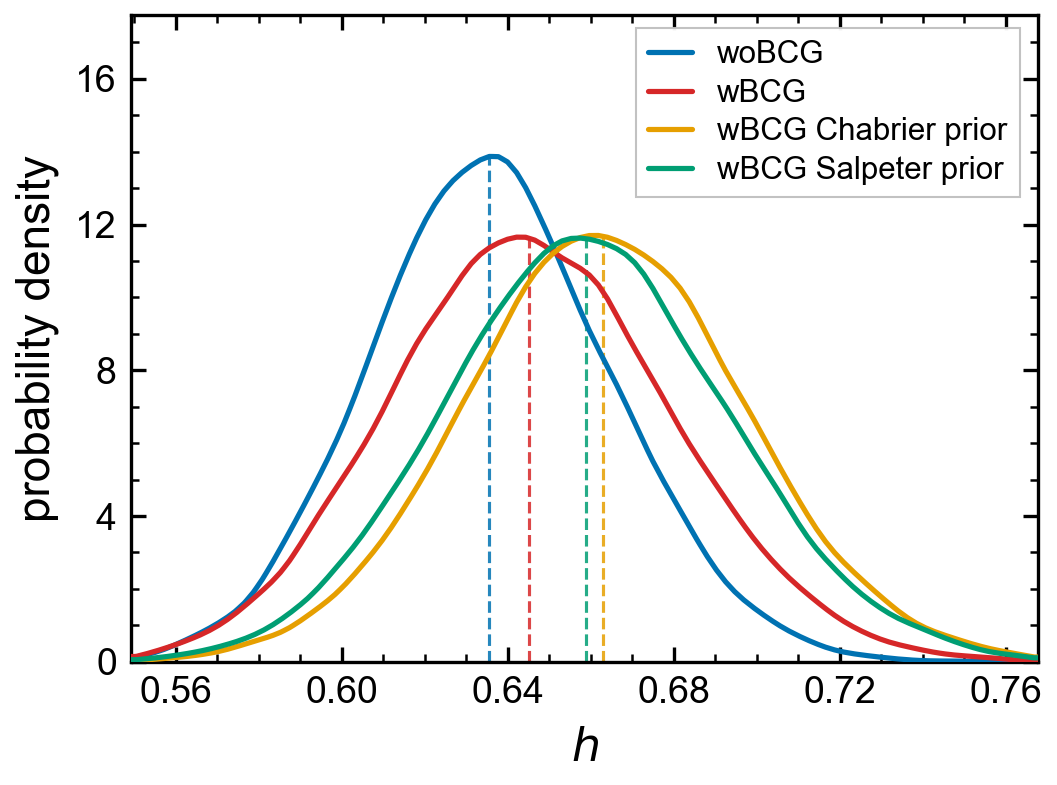}%
\label{fig:macsj1149_h_compare}}
\caption{Similar to Fig.~\ref{fig:sdss1004_compare}, but for MACS J1149.5+2223.}
\label{fig:macsj1149_compare}
\end{figure*}

We next consider the model in which the BCG is removed from 
the scaled galaxy population and represented by a separate Hernquist 
component, without imposing a stellar-mass prior on $M_{\rm BCG}$. 
The combined sample contains about $2.5\times10^{6}$ points and 
the minimum $\chi^2$ of the sampled models is $70.25$.
The marginalized posteriors are listed in the ``No prior'' column of
Table~\ref{tab:macsj1149}.  The main halo mass and concentration are
$M_1=(5.28^{+1.39}_{-1.21})\times10^{14}\,M_\odot/h$ and
$c_1=10.6^{+2.3}_{-3.3}$.  The core-size parameter is skewed toward large
values, $\beta=0.523^{+0.331}_{-0.350}$, or 
$\log_{10}\beta=-0.28^{+0.21}_{-0.48}$, corresponding to 
$r_{\rm c}\simeq60\,h^{-1}\,\mathrm{kpc}$.  The BCG mass is
$M_{\rm BCG}=(3.41^{+3.03}_{-1.61})\times10^{12}\,M_\odot/h$,
or $\log_{10}(M_{\rm BCG}/[M_\odot/h])=12.5^{+0.28}_{-0.28}$.
Figure~\ref{fig:macsj1149}(b) shows a positive correlation between
$\beta$ and $M_{\rm BCG}$, while neither parameter is strongly
correlated with $h$.  The marginalized $M_{\rm BCG}$ distribution is
skewed toward low values, with a tail extending to much larger masses.
The dimensionless Hubble parameter of the combined posterior is
$h=0.645^{+0.035}_{-0.034}$. Compared with the model without a 
separate BCG, both $\beta$ and $H_0$ are larger, as in 
the corresponding comparison for SDSS J1004+4112. 
While the MACS J1149.5+2223 posterior without a BCG mass prior
is more complex, the core-size parameter remains of order unity
and $M_{\rm BCG}$ is only weakly constrained toward high mass.

We then impose a Gaussian prior based on a Chabrier IMF stellar-mass
estimate on $M_{\rm BCG}$ of the same separate-BCG model,
$M_{\rm BCG}^{\rm Chabrier}=(3.146\pm0.815)\times10^{11}\,M_\odot/h$.
The combined sample contains about $2.3\times10^{6}$ and the minimum
$\chi^2$ of the sampled models is $71.80$. The marginalized 
posteriors are listed in the ``Chabrier'' column of
Table~\ref{tab:macsj1149}.  The main halo mass and concentration are
$M_1=(4.45^{+1.35}_{-1.15})\times10^{14}\,M_\odot/h$ and
$c_1=5.51^{+1.31}_{-0.74}$.  The core-size parameter is
$ \beta=0.0184^{+0.0337}_{-0.0140}$, 
or $\log_{10}\beta=-1.74^{+0.45}_{-0.62}$.  The lower edge of the
$68\%$ interval is consistent with a small core.  
The BCG mass is tightly constrained to 
$M_{\rm BCG}=(3.40^{+0.82}_{-0.83})\times10^{11}\,M_\odot/h$,
within $0.3\sigma$ of the Chabrier prior mean.  The dimensionless 
Hubble parameter is $h=0.663^{+0.034}_{-0.033}$.
Compared with the model without a BCG mass prior, the Chabrier prior
tightens $M_{\rm BCG}$ and reduces $\beta$ by more
than an order of magnitude, bringing the core size closer to that of
the model without a separate BCG. However, the inferred $H_0$ is
essentially unchanged.  Figure~\ref{fig:macsj1149}(c)
shows the joint posterior of $\log_{10}\beta$, $M_{\rm BCG}$,
and $h$.  There is a positive correlation between $\beta$ and
$M_{\rm BCG}$, while neither parameter is strongly correlated with $h$.

We repeat the separate-BCG analysis with a Gaussian prior based on a
Salpeter IMF stellar-mass estimate,
$M_{\rm BCG}^{\rm Salpter}=(4.856\pm0.830)\times10^{11}\,M_\odot/h$.
The combined sample contains about $2.3\times10^{6}$ points and 
the minimum $\chi^2$ of the sampled models is $71.82$.
The marginalized posteriors are listed in the ``Salpeter'' column of
Table~\ref{tab:macsj1149}.  The main halo mass and concentration are
$M_1=(4.26^{+1.39}_{-1.12})\times10^{14}\,M_\odot/h$ and
$c_1=6.13^{+2.22}_{-1.08}$.  The core-size parameter is
$\beta=0.0410^{+0.0865}_{-0.0276}$,
or $\log_{10}\beta=-1.39\pm0.49$.  The BCG mass is
$M_{\rm BCG}=(5.10\pm0.82)\times10^{11}\,M_\odot/h$,
within $0.3\sigma$ of the Salpeter prior mean.  The dimensionless 
Hubble parameter is $h=0.659^{+0.035}_{-0.034}$.
Compared with the Chabrier-prior model, the Salpeter prior allows a
larger BCG mass and a somewhat larger $\beta$, while $H_0$ is
essentially unchanged.  Both $\beta$ and $M_{\rm BCG}$ remain far
below the unconstrained separate-BCG solution obtained without a BCG
mass prior.  Figure~\ref{fig:macsj1149}(d) shows the
joint posterior of $\log_{10}\beta$, $M_{\rm BCG}$, and
$h$.  The positive correlation between $\beta$ and $M_{\rm BCG}$
persists, and neither parameter is strongly correlated with $h$.

The marginalized posteriors are compared directly in
Figure~\ref{fig:macsj1149_compare}.  Allowing the mass of the separate
BCG to vary without a stellar-mass prior shifts the core-size
distribution to substantially larger values,
$\log_{10}\beta=-0.28^{+0.21}_{-0.48}$.  The scaled-BCG result,
$-1.27^{+0.19}_{-0.17}$, is more tightly constrained, while the
Chabrier- and Salpeter-prior distributions are broader and overlap
the scaled-BCG distribution.  Their medians,
$\log_{10}\beta=-1.74^{+0.45}_{-0.62}$ and
$-1.39\pm0.49$, respectively, show that imposing either
stellar-mass prior returns the inference toward the small-core regime,
although the $68\%$ intervals still include both a nearly cuspy profile 
without a core and $\beta\simeq0.05$. 

In contrast, the four marginalized $h$ distributions overlap
substantially.  Their medians span only $0.635$--$0.663$, which is 
compared with individual $68\%$ credible half-widths of approximately
$0.03$--$0.04$.  The corresponding $H_0$ values,
$63.5^{+3.0}_{-2.8}\Mpc$ to $66.3^{+3.4}_{-3.3}\Mpc$, agree within
$1\sigma$ with one another and with the NFW-based measurement
$H_0=66.0\pm4.3\Mpc$ from SN Refsdal
\cite{2025PhRvD.112l3526L}, as well as with the earlier two-model
result $H_0=66.6^{+4.1}_{-3.3}\Mpc$ \cite{2023Sci...380.1322K}.
The Chabrier- and Salpeter-prior results are nearly
indistinguishable, and the no-prior result overlaps both the
scaled-BCG and stellar-mass-prior solutions.  The minimum $\chi^2$
values, $67.27$, $70.25$, $71.80$, and $71.82$, likewise differ by only
a few units, unlike the $\Delta\chi^2\simeq33$ induced by the same
prior in SDSS J1004+4112.  Thus, for MACS J1149.5+2223, 
the treatment of the BCG has a non-negligible effect
on the inferred core size but does not produce a statistically
resolved shift in $H_0$ at the precision of the present analysis.

\subsection{Implication for the self-interacting dark matter cross section}

Collisionless simulations predict a cuspy NFW profile
\cite{1997ApJ...490..493N}.  Shallower inner profiles have been
inferred in some relaxed clusters
\cite{2013ApJ...765...24N,2013ApJ...765...25N}, and strong-lensing
cores have been used to place upper limits on the self-interacting
dark matter (SIDM) cross section \cite{2022MNRAS.510...54A}.  Those
limits, and any SIDM reading of a large core
\cite{2026A&A...711A.275D}, assume that the stellar mass of the BCG
is independently constrained and is not free to absorb the same
central density.  In both of our systems a free Hernquist BCG
produces $\beta$ of order unity, whereas a Chabrier or Salpeter prior
returns $\beta\lesssim0.05$.  The same pattern was found for
Abell~611 after a stellar mass-to-light prior, where a mismatched
stellar component can also generate a spurious core
\cite{2019MNRAS.487.1905A}.  The large cores obtained here without a
BCG mass prior are therefore not evidence for self-interactions, but
they are degenerate with an unphysical stellar mass, and we do not
convert them into a cross section.  To be conservative, we restrict 
the conversion to the Salpeter-prior models, for which $M_{\rm BCG}$ 
is held to a stellar-population estimate.  

For the conversion, we adopt the scattering-rate condition
\cite{2016PhRvL.116d1302K,2022MNRAS.510...54A}
\begin{equation}
 \frac{\langle\sigma v\rangle}{m}\,\rho(r_{1})\,t_{\rm age}=1.
 \label{eq:sidm_rate}
\end{equation}
For a velocity-independent transfer cross section and a Maxwellian
velocity distribution, $\langle\sigma v\rangle=\sigma\langle v\rangle$
with
\begin{equation}
 \langle v\rangle=\frac{4}{\sqrt{\pi}}\,\sigma_{0},
\end{equation}
where $\sigma_{0}$ is the one-dimensional central dispersion.

Each sample of the main \texttt{acnfw} halo is converted as follows.
SDSS J1004+4112 has a single cluster-scale halo.  MACS J1149.5+2223
has four; we convert only the main \texttt{acnfw} halo, not the
three additional NFW components.   The virial radius is
computed from the sampled $M$ and $h$ at the cluster redshift, using
the overdensity of Ref.~\cite{1998ApJ...495...80B} and the
cosmology of that {\sc glafic} input.  Then
$r_{\rm s}=r_{\rm vir}/c$, $r_{\rm c}=\beta r_{\rm s}$, and $\rho_{s}$
follows from the NFW mass integral.  The
circular-velocity maximum $V_{\rm max}$ is that of the corresponding
NFW halo.  Ref.~\cite{2022MNRAS.510...54A} 
infer $\sigma_{0}$ from the isothermal
Jeans equation and their Table~4 gives
$\sigma_{0}/V_{\rm max}\simeq0.45$--$0.59$ (median $0.54$).  We set
$\sigma_{0}=0.54\,V_{\rm max}$ rather than $V_{\rm max}/\sqrt{2}$,
which would overestimate $\langle v\rangle$.  The halo age
$t_{\rm age}$ is the age of the Universe at the lens redshift for
$\Omega_{\mathrm{m}}=0.3$ and $h=0.7$, independent of the time-delay $h$
inferred in the lens model.

In the SIDM model, $r_{1}$ is the
matching radius between the isothermal core and the outer NFW
envelope, which differs from the cored NFW parameter $r_{\rm c}$.  
For the three clusters with well-measured cores
($r_{\rm c}\gtrsim15\,\mathrm{kpc}$) 
in Ref.~\cite{2022MNRAS.510...54A}, the ratios are
$r_{1}/r_{\rm c}=5.9$, $5.8$, and $2.8$. We thus take the median value, 
$r_{1}=5.8\,r_{\rm c}$, to evaluate $\rho(r_{1})$ on the cored NFW
profile of each sample.  Since our Salpeter-prior cores are smaller than
that calibration sample, the factor $5.8$ is an extrapolation
rather than a new isothermal fit to $\Sigma(R)$.

With these assumptions, we derive medians and equal-tailed $68\%$ intervals
by transforming the joint posterior of $(\beta,M,c,h)$
sample by sample.  
For SDSS J1004+4112, the Salpeter-prior $\beta$ posterior
piles up at $\sim10^{-3}$ (Table~\ref{tab:sdss1004}) and is an upper
limit.  The main halo has
$M_{\rm vir}=(11.5^{+1.2}_{-1.1})\times10^{14}\,M_{\odot}$,
$r_{\rm vir}=2285^{+141}_{-138}\,\mathrm{kpc}$, and
$r_{\rm s}=524^{+56}_{-55}\,\mathrm{kpc}$.  The $95\%$ one-sided
credible bound is $\beta<1.12\times10^{-2}$, or
$r_{\rm c}<5.3\,\mathrm{kpc}$, and the $68\%$ interval is
$r_{\rm c}=1.16^{+1.94}_{-0.85}\,\mathrm{kpc}$.  From
$V_{\rm max}=1533^{+29}_{-28}\,\kms$, we take
$\sigma_{0}=828^{+16}_{-15}\,\kms$ and
$\langle v\rangle=1868^{+35}_{-34}\,\kms$.  With
$r_{1}=6.7^{+11.3}_{-4.9}\,\mathrm{kpc}$ and
$t_{\rm age}=7.10\,\mathrm{Gyr}$, Eq.~(\ref{eq:sidm_rate}) gives
$\sigma/m<0.035\,\mathrm{cm}^{2}\,\mathrm{g}^{-1}$ ($95\%$,
one-sided).  The Chabrier-prior model of the same cluster has a
still smaller $84\%$ quantile in $\beta$ and would only tighten that
bound.

For MACS J1149.5+2223, the Salpeter-prior posterior is
peaked at a finite core,
$\beta=0.0410^{+0.0865}_{-0.0276}$ ($68\%$;
Table~\ref{tab:macsj1149}), but the interval still includes a nearly
cuspy profile.   The main halo has
$M_{\rm vir}=(6.46^{+2.20}_{-1.74})\times10^{14}\,M_{\odot}$,
$r_{\rm vir}=1721^{+205}_{-191}\,\mathrm{kpc}$,
$r_{\rm s}=280^{+93}_{-94}\,\mathrm{kpc}$, and
$r_{\rm c}=11.7^{+13.3}_{-6.9}\,\mathrm{kpc}$.  From
$V_{\rm max}=1409^{+95}_{-91}\,\kms$, we take
$\sigma_{0}=761^{+51}_{-49}\,\kms$ and
$\langle v\rangle=1718^{+115}_{-111}\,\kms$.  With
$r_{1}=68^{+77}_{-40}\,\mathrm{kpc}$ and
$t_{\rm age}=7.97\,\mathrm{Gyr}$, Eq.~(\ref{eq:sidm_rate}) gives
$\sigma/m=0.071^{+0.15}_{-0.045}\,\mathrm{cm}^{2}\,\mathrm{g}^{-1}$
 ($68\%$ credible interval). The equal-tailed $95\%$ interval is
$[0.0035,\,1.58]\,\mathrm{cm}^{2}\,\mathrm{g}^{-1}$.
That interval is not recast as a one-sided upper limit, 
as the posterior is not piled up at very small $\beta$.

The two translations are therefore not a common measurement of
$\sigma/m$.  SDSS J1004+4112 is a cusp-like upper limit;
MACS J1149.5+2223 remains consistent with that limit and with a
modest core. We note that here we just convert the constraints on 
the core size into the SIDM cross section ignoring any possible 
effects of the baryon physics, and when non-zero $\beta$ can be
produced by the baryon physics the constraint on the SIDM cross 
section from MACS J1149.5+2223 should be modified.

\section{Conclusion}\label{sec:conclusions}

We have investigated the effect of a finite core of the central 
dark matter density profile on the inference of the Hubble constant
with cluster strong lensing. Adopting a cored elliptical NFW density 
profile, we have studied the effect of the core in detail with 
two cluster strong lens systems, SDSS J1004+4112 and MACS J1149.5+2223, for
which time delays have been measured for multiply imaged quasars and
supernovae, respectively. Since the central dark matter distribution
is expected to degenerate with the mass distribution of the BCG, we 
consider four different treatments of the BCG.

For SDSS J1004+4112, the unconstrained separate-BCG model gives
a large core, $\beta\simeq0.5$, and
$H_0=67.2^{+4.8}_{-4.5}\Mpc$.  It attains the lowest $\chi^2$, but
only by driving $M_{\rm BCG}$ two orders of magnitude above
stellar-population estimates. Once either IMF prior is imposed, 
the $\beta$ posterior piles up near $10^{-3}$, while $H_0$ falls 
to $53$--$54\Mpc$. For instance, with the Salpeter prior we have 
obtained $\beta=(2.18^{+4.03}_{-1.60})\times10^{-3}$ and
$H_0=54.4^{+2.7}_{-2.5}\Mpc$. The scaled-BCG model occupies the same 
small-core regime and gives an intermediate $H_0=56.9\pm2.8\Mpc$.

MACS J1149.5+2223 does not follow the same trend as SDSS J1004+4112.  
The scaled-BCG posterior of the core-size parameter 
is already peaked at $\beta\simeq0.05$.  
A free BCG mass again produces $\beta$ of order unity, and the 
IMF priors return $\beta\simeq0.02$--$0.04$.  The four models fit
comparably well, and $H_0$ remains $63$--$66\Mpc$.  
As a representative example, with the Salpeter prior we have 
obtained $\beta=0.0410^{+0.0865}_{-0.0276}$ and $H_0=65.9^{+3.5}_{-3.4}\Mpc$. 

Because $\beta$ and $h$ are only weakly correlated within each 
sampled posterior, the difference between the two clusters is not 
a one-parameter degeneracy
with the core size, but rather a change in the central mass 
decomposition, and the cosmographic response of that change is 
system-dependent.

In both of the cluster lenses with well-measured time delays, a finite
dark matter core and the stellar mass of the BCG have to be varied
together.  Replacing the main halo with a cored NFW profile and
leaving the BCG mass free yield the core-size parameter 
$\beta\sim\mathcal{O}(1)$.  A Chabrier or Salpeter Gaussian prior 
on $M_{\rm BCG}$ suppresses the core size.  The corresponding shift 
in $H_0$ is not universal, it exceeds $10\Mpc$ in 
SDSS J1004+4112 and is unresolved in MACS J1149.5+2223. 
This difference highlights the critical importance of having many 
multiple images to robustly constrain the lens mass distribution, 
which was also suggested by the previous analysis of the dependence of
time-delay cosmographic results with galaxy clusters on mass-model choices 
\citep{2023PhRvD.108h3532L,2025PhRvD.111l3506L}.

We have also translated the constraints on the core size into 
those of the SIDM cross section. Under a stellar-mass prior 
assuming the Salpeter IMF,  we have obtained the 95\% upper limit 
of $\sigma/m<0.035\,\mathrm{cm}^{2}\,\mathrm{g}^{-1}$ from 
SDSS J1004+4112, and the 68\% range of $\sigma/m=0.071^{+0.15}_{-0.045}\,\mathrm{cm}^{2}\,\mathrm{g}^{-1}$
from MACS J1149.5+2223. Our result is consistent with 
Ref.~\cite{2022MNRAS.510...54A}.

\begin{acknowledgments}
Y.L. was supported by the Natural
Science Foundation of Inner Mongolia of China (Grant No.2025QN01042) and Research Start-up Fund of Inner Mongolia University (Grant No.10000-A25201021). M.O. was supported by JSPS KAKENHI Grants Numbers JP25H00662, JP25H00672, and JP22K21349.  
\end{acknowledgments}

\appendix

\section{Cored Elliptical NFW profile}\label{app:cnfw}

The cored NFW density profile, which is an extension of the
Navarro-Frenk-White density profile \cite{1997ApJ...490..493N} with a
finite core and is introduced in previous strong lensing studied
\cite{2013ApJ...765...24N,2013ApJ...765...25N,2019MNRAS.487.1905A,2022MNRAS.510...54A},
has the following form of the three-dimensional density profile
\begin{equation}
  \rho(r)=\frac{\rho_{\mathrm{s}}}{(r_{\mathrm{c}}/r_{\mathrm{s}}+r/r_{\mathrm{s}})(1+r/r_{\mathrm{s}})^2},
\end{equation}
where $\rho_{\mathrm{s}}$ is the characteristic density,
$r_{\mathrm{s}}$ is the scale radius, and $r_{\mathrm{c}}$ is the core
radius. In addition to the concentration parameter, it is convenient
to define the following dimensionless parameter
\begin{equation}
  \beta=\frac{r_{\mathrm{c}}}{r_{\mathrm{s}}},
\end{equation}
which quantifies the core radius. The limit $\beta\rightarrow 0$
corresponds to the standard NFW profile. Analytic expressions of the
convergence and the deflection angle of the spherically symmetric
cored NFW profile are presented in Ref.~\cite{2022MNRAS.510...54A}.

For strong lensing analyses we very often use elliptical extensions of
various density profiles. While the numerical integration is used to
computed various lensing properties of the cored elliptical NFW
profile in Ref.~\cite{2022MNRAS.510...54A}, here we adopt an
alternative approach to model the cored elliptical NFW profile by a
superposition of cored steep ellipsoids, which enables fast
calculations of lensing properties, following
Ref.~\cite{2021PASP..133g4504O}. A challenge is that in this case the
fitting parameters are not constant but should be written as a
function of $\beta$. To do so, we change $\log_{10}\beta$ in the range
between $-8$ and $2$ with an interval of $0.05$ to derive 88 fitting
parameters, which specifies normalizations and core
radii of individual cored steep ellipsoids, for each $\beta$.
We then tabulate the 88 fitting parameters for all the 201
$\beta$ values, and derive the fitting parameters for each input
$\beta$ simply by a linear interpolation on $\beta$. This profile is
already implement in a public version of {\sc glafic} as {\texttt{acnfw}}.

\section{Stellar masses of BCGs from the photometry}\label{app:bcg_mass}

\subsection{SDSS J1004+4112}\label{app:bcg_mass_1004}

We take the HST photometry of the BCG (CASTLES object G) from the
CASTLES database \cite{2001ASPC..237...25F,KochanekCASTLES}. The quasar images
A--E are not used. The adopted Vega magnitudes are
$m_{\mathrm{F555W}}=23.22\pm0.50$,
$m_{\mathrm{F814W}}=18.84\pm0.18$, and
$m_{\mathrm{F160W}}=17.23\pm0.32$.
Since CASTLES lists no uncertainty on F555W, we replace it with
$0.50$\,mag. Magnitudes are converted to AB magnitudes using 
instrument-specific AB$-$Vega offsets, $-0.010$ for ACS/WFC F555W 
and $+0.431$ for ACS/WFC F814W (STScI ACS zeropoint calculator), 
and $+1.315$ for NICMOS/NIC2 F160W \cite{2015AJ....149..159R}. 
Flux densities follow $F_{\nu}/\mathrm{mJy}=3631\times10^{-0.4m_{\mathrm{AB}}}$.
The resulting fluxes, in mJy, are
$0.00189\pm0.00087$ (F555W),
$0.0711\pm0.0118$ (F814W) and
$0.1387\pm0.0409$ (F160W).
The redshift is fixed at $z_{\rm l}=0.68$.

Stellar masses are estimated with \textsc{cigale} \cite{2019A&A...622A.103B}.
The models use a delayed star-formation history with no late burst, the
\citet{2003MNRAS.344.1000B} simple stellar populations, a modified Calzetti
attenuation law \cite{2000ApJ...533..682C}, and the CIGALE redshifting 
module. The grid is
$\tau_{\mathrm{main}}=\{0.5,1,2,4\}$\,Gyr,
$t_{\mathrm{main}}=\{3,5,7\}$\,Gyr (all younger than the age of the
Universe at $z=0.68$),
$Z=\{0.02,0.05\}$ and
$E(B-V)_{\rm l}=\{0.0,0.1,0.2\}$.
The F555W, F814W and F160W fluxes are fitted with the ACS/WFC F555W,
ACS/WFC F814W and NICMOS/NIC2 F160W filter curves in CIGALE.
We run the same grid twice, changing only the IMF. CIGALE adopts a
Planck 2018 cosmology \cite{2020A&A...641A...6P}, and the luminosity distance
returned by the fit is $D_{L}=4233$\,Mpc, corresponding to
$H_{0}\simeq67.7\,\mathrm{km\,s^{-1}\,Mpc^{-1}}$.

The Bayesian stellar masses are
$M_{\star}=(3.46\pm0.85)\times10^{11}\,M_{\odot}$
for a Chabrier IMF ($\chi^{2}_{\nu}=4.41$) and
$M_{\star}=(6.09\pm1.49)\times10^{11}\,M_{\odot}$
for a Salpeter IMF ($\chi^{2}_{\nu}=4.49$).
The two fits have essentially the same $\chi^{2}$. The mass ratio is
$1.76$ ($0.25$\,dex), as expected from the IMF normalization.
The preferred models are old ($t_{\mathrm{main}}\sim6$\,Gyr),
metal-rich and lightly reddened.

These values are the stellar masses associated with the CASTLES fluxes.
They are not infinite-aperture integrals. If the CASTLES photometry
misses outer envelope or intracluster light, $M_{\star}$ underestimates
the Hernquist total mass $M_{\mathrm{tot}}$. With only three bands and
$\chi^{2}_{\nu}\simeq4.4$, the formal CIGALE errors also underestimate 
the true uncertainty. We therefore do not use the statistical error as the
prior width. Scaling that error by $\sqrt{\chi^{2}_{\nu}}$ gives
$\sim0.22$\,dex. Allowing in addition for aperture definition, the BCG/ICL
separation and residual contamination, we adopt $0.25$\,dex as a more 
conservative error.

\textsc{glafic} \cite{2010PASJ...62.1017O} takes the Hernquist mass 
in units of $M_{\odot}/h$. Assuming the fiducial value of the Hubble
constant of $h=0.7$, the stellar masses are converted to 
$2.27\times10^{11}\,M_{\odot}/h$ (Chabrier) and
$3.99\times10^{11}\,M_{\odot}/h$ (Salpeter). With the 
$0.25$\,dex error, the priors used in \textsc{glafic} are
\begin{align}
  M_{\mathrm{BCG}}^{\mathrm{Chabrier}}
  &= (2.27\pm1.31)\times10^{11}\,M_{\odot}/h\,,\\
  M_{\mathrm{BCG}}^{\mathrm{Salpeter}}
  &= (3.99\pm2.30)\times10^{11}\,M_{\odot}/h\,.
\end{align}
The two IMFs are not
combined in a single prior, but each is used as a separate lens-model run
and the difference is treated as a systematic.

\subsection{MACS J1149.5+2223}\label{app:bcg_mass_1149}

The photometry is taken from the CLASH ACS+IR catalog
\cite{2012ApJS..199...25P}, source ID 5251. The measurements are
isophotal, with a common isophote of area $36560$~pixels
($0.065$\,arcsec pixel$^{-1}$), or $154.5$\,arcsec$^{2}$. The equivalent
circular radius is $7.0$\,arcsec, which is $45$\,kpc at $z=0.541$ for
$h=0.7$. Since the Galactic extinction has already been removed in the catalog
($E(B-V)=0.02297$), we do not apply a further correction. 
The redshift is fixed at the GLASS spectroscopic value $z=0.541$
\cite{2015ApJ...812..114T}.

The eleven HST AB magnitudes are
F435W$=20.656$, F555W$=19.875$, F606W$=19.258$, F625W$=18.976$,
F775W$=18.168$, F814W$=18.122$, F105W$=17.564$, F110W$=17.457$,
F125W$=17.319$, F140W$=17.164$ and F160W$=17.013$.
Statistical errors on this source in the catalog are as small as
$0.0006$\,mag in F160W. Those uncertainties describe the photon noise rather 
than the aperture definition, the PSF matching or the subtraction 
of an intracluster light.
We therefore impose a floor of $0.05$\,mag in every HST band before
the fit. Flux densities follow
$F_{\nu}/\mathrm{mJy}=3631\times10^{-0.4m_{\mathrm{AB}}}$.

The CLASH photometry is formally isophotal. For this BCG it is consistent
with a total-light measurement. In the ASTRODEEP catalog the BCG is
source ID 100017, a bright cluster member whose HST magnitude is the
total magnitude of a Galfit analytic profile, with ICL fitted separately
\cite{2016A&A...590A..30M,2017A&A...607A..30D}. The F160W magnitudes
are $17.013$ (CLASH) and $17.048$ (ASTRODEEP), a difference of
$0.035$\,mag ($\sim 3$ per cent in flux). We therefore treat the
stellar mass derived from the CLASH spectral energy distribution (SED) 
as a total stellar mass and do
not apply an extra aperture correction. While a de Vaucouleurs extrapolation
with $R_{e}=3$--$8$\,arcsec would add $0.14$--$0.33$\,dex, this 
correction is inconsistent with the observed total magnitude and therefore 
is not used.

ASTRODEEP also reports Ks and IRAC 3.6 and $4.5\,\mu\mathrm{m}$ fluxes
from T-PHOT, using the Galfit profile as a spatial prior on the
lower-resolution images \cite{2017A&A...607A..30D}. Adding those three
bands changes the Chabrier mass by only $+0.015$\,dex, but
$\chi^{2}_{\nu}$ rises from $0.32$ to $1.37$. The increase is confined
to $K_s$ (residual $\approx-2.4\sigma$, the observed flux fainter than the
model) and IRAC2 ($\approx+3.0\sigma$, the observed flux brighter than the
model), which pull in opposite directions. The eleven HST bands remain
within $\sim1.5\sigma$. Since the T-PHOT photometry of a BCG in a 
crowded, ICL-contaminated core is not reliable at that level, 
the adopted prior uses the HST bands only.

Stellar masses are also estimated with \textsc{cigale}
\cite{2019A&A...622A.103B}. The models use a delayed star-formation
history, the \citet{2003MNRAS.344.1000B} simple stellar populations, the
CIGALE modified-starburst attenuation law \cite{2000ApJ...533..682C}
and the CIGALE redshifting module. The grid is
$\tau_{\mathrm{main}}=\{0.1,0.3,0.5,1,2,4,8\}$\,Gyr,
$t_{\mathrm{main}}=\{2,3,4,5,6,7,7.5\}$\,Gyr (all younger than the age of
the Universe at $z=0.541$),
$f_{\mathrm{burst}}=\{0,0.01,0.05\}$,
$Z=\{0.008,0.02,0.05\}$ and
$E(B-V)_{\ell}=\{0.0,0.1,0.2,0.3,0.5\}$, with
$E(B-V)_{\mathrm{s}}=0.44\,E(B-V)_{\ell}$.
The eleven fluxes are fitted with the ACS/WFC and WFC3/IR filter curves
in CIGALE. We run the same grid twice, changing only the IMF.

The Bayesian estimates of the stellar mass are
$\log_{10}(M_{\star}/M_{\odot})=11.679\pm0.112$
($\chi^{2}_{\nu}=0.32$) for a Chabrier IMF and
$\log_{10}(M_{\star}/M_{\odot})=11.867\pm0.074$
($\chi^{2}_{\nu}=0.31$) for a Salpeter IMF,
or
$M_{\star}=(4.77\pm1.23)\times10^{11}\,M_{\odot}$ and
$(7.36\pm1.26)\times10^{11}\,M_{\odot}$
in the CIGALE cosmology. The mass ratio is $1.54$ ($0.19$\,dex), which is
smaller than the $0.24$\,dex shift from IMF normalization alone, because
the best-fitting age and attenuation also change. Specifically 
the Salpeter solution is younger and more reddened. 
The reduced $\chi^{2}$ values are below unity because of the 
$0.05$\,mag error floor, not because the photometry is overfitted 
in a useful sense. Mock catalogs recover the input Chabrier 
mass to $0.1\sigma$.

We follow the same procedure converting from the CIGALE cosmology 
and from physical solar masses to \textsc{glafic} units to obtain 
$3.146\times10^{11}\,M_{\odot}/h$ (Chabrier) and
$4.856\times10^{11}\,M_{\odot}/h$ (Salpeter).

Each prior width is the CIGALE Bayesian uncertainty on
$\log_{10}M_{\star}$ for that IMF, converted to a linear Gaussian,
$\sigma_{M}=(\ln 10)\,\sigma_{\mathrm{dex}}\,M$, with
$\sigma_{\mathrm{dex}}=0.112$ (Chabrier) and $0.074$ (Salpeter).
The Gaussian therefore describes only the statistical posterior of the
SED fit. We do not multiply $\sigma_{\mathrm{dex}}$ by
$\sqrt{\chi^{2}_{\nu}}$, which is sometimes used when
$\chi^{2}_{\nu}>1$ to correct an underestimated error. Here
$\chi^{2}_{\nu}\simeq0.32$ because of the $0.05$\,mag floor, so the
factor would make the prior narrower, not more conservative.
The priors used in \textsc{glafic} are
\begin{align}
  M_{\mathrm{BCG}}^{\mathrm{Chabrier}}
  &= (3.146\pm0.815)\times10^{11}\,M_{\odot}/h\,,\\
  M_{\mathrm{BCG}}^{\mathrm{Salpeter}}
  &= (4.856\pm0.830)\times10^{11}\,M_{\odot}/h\,.
\end{align}
Again, the two IMFs are not
combined in a single prior, but each is used as a separate lens-model run
and the difference is treated as a systematic.

\bibliography{ref}

\end{document}